\documentclass[prl,twocolumn,notitlepage,superscriptaddress,showkeys,showpacs]{revtex4-1}
\usepackage{amsmath,amsfonts,amssymb,amsthm,epsfig,array}
\usepackage{dsfont}
\usepackage{ulem}
\usepackage{graphics}
\usepackage{float}
\usepackage{bm}
\usepackage{verbatim}
\usepackage[dvipsnames]{xcolor}
\usepackage[hidelinks,colorlinks,linkcolor=blue,
citecolor=blue,urlcolor=blue]{hyperref}
\usepackage[titletoc,title]{appendix}

\let\oldAA\AA
\renewcommand{\AA}{\text{\normalfont\oldAA}}

\date{\today}

\begin{document}
\title{High-$T_c$ Superconductor Fe(Se,Te) Monolayer: an Intrinsic, Scalable and Electrically-tunable Majorana Platform}
\author{Xianxin Wu}
\affiliation{Institut f\"ur Theoretische Physik und Astrophysik,
  Julius-Maximilians-Universit\"at W\"urzburg, 97074 W\"urzburg,
  Germany}
\author{Xin Liu}
\affiliation{School of Physics, Huazhong University of Science and Technology, Wuhan, Hubei 430074, China}
\affiliation{Wuhan National High Magnetic Field Center, Huazhong University of Science and Technology, Wuhan, Hubei 430074, China}
\author{Ronny Thomale}
\affiliation{Institut f\"ur Theoretische Physik und Astrophysik,
  Julius-Maximilians-Universit\"at W\"urzburg, 97074 W\"urzburg,
  Germany}
\author{Chao-Xing Liu}
\email{cxl56@psu.edu}
\affiliation{Department of Physics, the Pennsylvania State University, University Park, PA, 16802}
\begin{abstract}
A monolayer of the high-$T_c$ superconductor FeTe$_{1-x}$Se$_x$ has been predicted to realize a topologically non-trivial state with helical edge modes at its boundary, providing a novel intrinsic system to search for topological superconductivity and Majorana zero modes. Evidence in favor of a topological phase transition and helical edge modes has been identified in recent experiments~\cite{Peng2019}. We propose to create Majorana zero modes by applying an in-plane magnetic field to the FeTe$_{1-x}$Se$_x$ monolayer and by tuning the local chemical potential via electric gating. Owing to the anisotropic magnetic couplings on edges from a topological band inversion, an in-plane magnetic field drives the system into an intrinsic high-order topological superconductor phase with Majorana corner modes, without fabricating heterostructures. Furthermore, we demonstrate that Majorana zero modes can occur at other different locations, including the domain wall of chemical potentials at one edge and certain type of tri-junction in the 2D bulk. Our study not only demonstrates FeTe$_{1-x}$Se$_x$ monolayer as a promising Majorana platform with scalability and electrical tunability and within reach of contemporary experimental capability, but also provides a general principle to search for realistic realization of
high-order topological superconductivity.
\end{abstract}
\maketitle


{\it Introduction: }
Within the nomenclature of condensed matter, a Majorana zero mode (MZM) is an anyonic quasi-particle excitation with non-Abelian statistics, which underpins the concept of topological quantum computations \cite{Ivanov2001,Kitaev2003,Kitaev2006,Nayak2008,Alicea2012,Sarma2015,Aasen2016,Karzig2017}. 
A variety of physical systems have been theoretically proposed to realize MZMs, including the $\nu=5/2$ fractional quantum Hall state\cite{Moore1991,Read2000,Nayak2008}, a chiral $p$-wave state possibly realized in Sr$_2$RuO$_4$ superconductors (SCs)\cite{RiceTM1995,DasSarma2006}, Pfaffian spin liquids~\cite{PhysRevLett.102.207203}, semiconducting nano-wires in proximity to SCs subject to magnetic fields \cite{Kitaev2001PU,Lutchyn2010,PhysRevLett.105.177002,Sau2010,AliceaJ2010}, the surface of topological insulators (TIs) in proximity to SCs\cite{Fu2008}, quantum anomalous Hall insulator-SC heterostructures\cite{Qi2010,Chung2011}, and ferromagnetic atomic chains on SCs\cite{Nadj-Perge2013PRB,Braunecker2013}. Major experimental efforts currently focus on heterostructures made of SCs and spin-orbit coupled systems (such as TIs or semiconducting nano-wires), in which evidences of MZMs have been found \cite{Mourik2012,WangMX2012,Nadj-Perge2014,Sun2016,Deng2016,ZhangH2018,Lutchyn2018}. Unambiguous detection and manipulation of MZMs in these heterostructures, however, heavily rely on the SC proximity effect that suffers from the complexity of the interface. Furthermore, the low operation temperature of conventional superconducting materials complicates further manipulation of MZMs.
It is thus desirable to find an intrinsic, robust, and controllable Majorana platform that is compatible with existing fabrication and patterning technologies. To this end, recent theoretical predictions and the experimental verification of a TSC phase at the surface of Fe(Se,Te) SCs \cite{Hao2014,Wu2016,Wang2015,XuPRL2016,Zhang2018,Zhang2019NP,WangDF2018} provides an exciting opportunity due to their intrinsic nature of both superconductivity and non-trivial band structure which further comes along with a comparably high critical temperature T$_c$. More recently, the direct observation of band inversion in two-dimensional (2D) Fe(Se,Te) monolayer suggests the coexistence of a quantum spin Hall (QSH) state and superconductivity, thus providing a new two-dimensional (2D) platform for MZMs \cite{Shi2017,Peng2019}, with a T$_c$ of 40 K \cite{Li2015PRB} and a large in-plane upper critical field of about 45 T\cite{Salamon2016}.

\begin{figure}[t]
\centerline{\includegraphics[width=1\columnwidth]{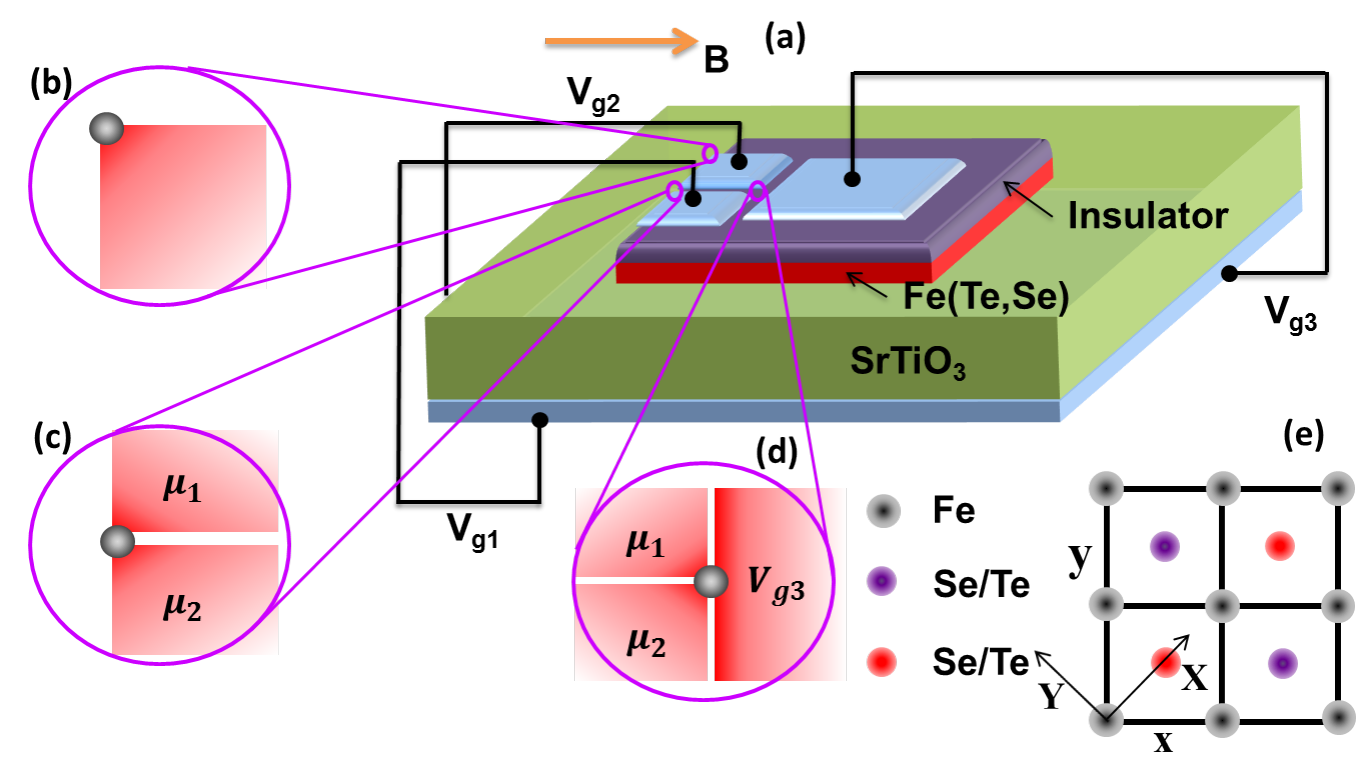}}
\caption{ Schematics for the Majorana platform based on Fe(Se,Te) monolayer. MZMs can be found at three different locations: (b) the corner between two perpendicular edges; (c) the CPDW along the 1D edge; (d) the tri-junction in the 2D bulk. $\mu_{1,2}$ label chemical potentials and $V_{g3}$ is for the asymmetric potential, both of which can be generated by a dual gate voltage.
The gray circles in (b), (c) and (d) represent the MZMs.
(e) The crystal structure for Fe(Te,Se) monolayer and the coordinate system. The gray circles are for Fe atoms and the red (purple) circles are for the Se/Te atoms above (below) the Fe layer. \label{setup} }
\end{figure}

In this work, we theoretically explore different feasible experimental configurations to realize MZMs in a Fe(Se,Te) monolayer by controlling the local chemical potential and the in-plane magnetic field. The experimental setup for a Fe(Te,Se) monolayer with local gating is shown in Fig.\ref{setup}(a). By studying the topological phase transition (TPT) at the one-dimensional (1D) edge and its dependence on the magnetic field direction, we demonstrate the existence of the MZMs at the corner of two perpendicular edges with the in-plane magnetic field parallel to one edge (Fig.\ref{setup}(b)), derived from a magnetic-anisotropy induced high-order topological superconductor phase,  and the chemical potential domain wall (CPDW) along 1D edge (Fig.\ref{setup}(c)). The magnetic anisotropy intrinsically originates from the topological band inversion between states at $\Gamma$ point with different total angular momenta of the QSH state. We further reveal a 2D bulk TPT between the QSH state and a trivial insulator induced by electric gating in the Fe(Te,Se) monolayer, due to which the MZM can also be trapped in a tri-junction (Fig.\ref{setup}(d)).

\textit{ Model Hamiltonian and TPT at the 1D edge }
We will first demonstrate the existence of TPT at the 1D edge in a Fe(Te,Se) monolayer upon applying an in-plane magnetic field. We start from a tight-binding model including five Fe 3$d$ orbitals and spin-orbit coupling (SOC) \cite{Wu2016}
\begin{eqnarray}
\mathcal{H}_0&=&\sum_{\alpha\beta}\sum_{mn\sigma}\sum_{ij}(t^{mn}_{\alpha\beta,ij}+\epsilon_{\alpha}\delta_{mn}\delta_{\alpha\beta}\delta_{ij})
c^{\dag}_{\alpha m\sigma}(i)c_{\beta n \sigma}(j)\nonumber\\
&&+\sum_{i\alpha\sigma\sigma'} \lambda_{soc} M^{\sigma\sigma'}_{mn} c^{\dag}_{\alpha m\sigma}(i)c_{\alpha n \sigma'}(i),
\end{eqnarray}
where $\alpha$,$\beta=A,B$ labels the sublattices (two Fe atoms in one unit-cell in Fig.\ref{setup}(e)), $\sigma$ labels spin, $m$,$n$ label five $d$ orbitals, and $i$,$j$ label the indices of unit-cell. $t^{mn}_{\alpha\beta,ij}$ are the hopping parameters, $\epsilon_{m}$ are the on-site energies of Fe $d$ orbitals and $\lambda_{soc}$ labels the SOC strength. The values of these parameters can be found in the Supplementary Materials (SM). $c^{\dag}_{\alpha m \sigma}(i)$ is the creation operator for an electron with spin-$\sigma$ and orbital-$m$ at the $\alpha$ sublattice site of the unit-cell $i$.
In order to treat superconductivity in a Fe(Se,Te) monolayer without entering a detailed microscopic derivation, we consider spin singlet intra-orbital pairing within the same sublattice, for which the SC Hamiltonian reads
\begin{eqnarray}
\mathcal{H}_{SC}=\sum_{\alpha m\sigma,ij}\sigma(\Delta_0\delta_{ij}+\frac{\Delta_1}{4}\delta_{\langle\langle ij \rangle\rangle})c^{\dag}_{\alpha m\sigma}(i)c^\dag_{\alpha m \bar{\sigma}}(j)+h.c.,\nonumber\\
\end{eqnarray}
where $\Delta_0$ is the on-site pairing and $\langle\langle ij \rangle\rangle$ labels the next nearest neighbor sites for the pairing parameter $\Delta_1$, which generates the well-known $s_{\pm}$-wave pairing in iron based superconductors\cite{Hirschfeld2011}. Here, we neglect the inter-orbital pairing, which may also exist \cite{Nourafkan2016}, but will not have any qualitative effect.
The Zeeman coupling is given by
\begin{eqnarray}
\mathcal{H}_{Z}\!=\!\!\!\!\!\!\sum_{\alpha mn\sigma\sigma'i}\!\!\!\mu_B g_0 \bm{B}\!\cdot\!(\bm{s}_{\sigma\sigma'}\delta_{mn}+\frac{1}{2}\bm{L}_{mn}\delta_{\sigma\sigma'} )c^{\dag}_{\alpha m\sigma}(i)c_{\alpha n \sigma'}(i)\nonumber\\
\end{eqnarray}
with the Bohr magneton $\mu_B$ and the g-factor $g_0$. Here $\bm{s}$ and $\bm{L}$ are spin and orbital angular momentum operators with the forms given in SM.

\begin{figure}[tb]
\centerline{\includegraphics[width=1\columnwidth]{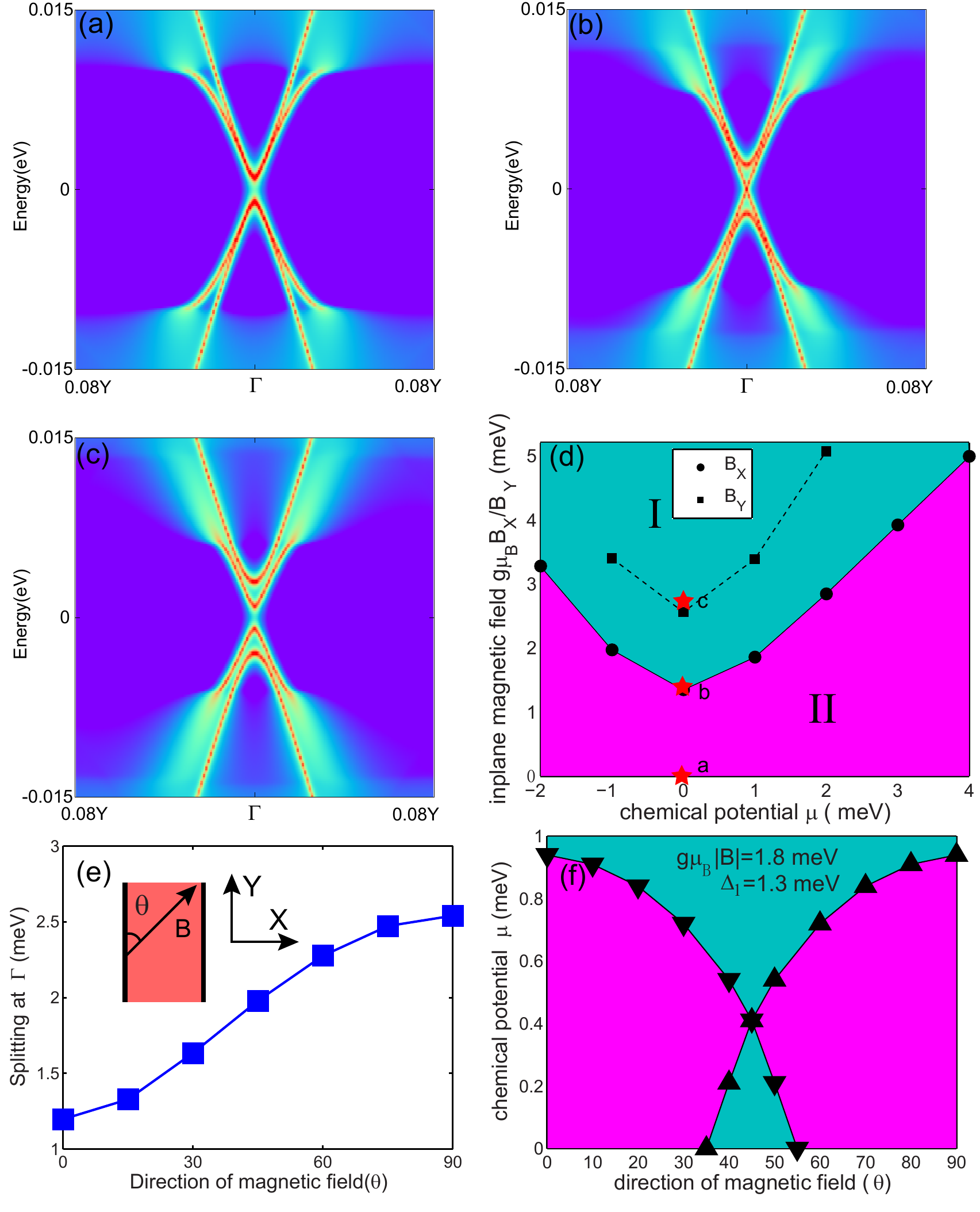}}
\caption{ TPT on edges and anisotropic magnetic coupling. Energy spectra of the semi-infinite system with the (100) edge as a function of the in-plane magnetic fields: (a) $B_X=0$, (b) $B_X=1.4$ meV and (c) $B_X=2.8$ meV. (d) TPT as a function of magnetic fields and chemical potentials for the (100) edge. The black circles (squares) correspond to the TPT line for the magnetic field $B_X$ ($B_Y$) perpendicular (parallel) to the edge.  The blue and pink regions represent the phases I and II, respectively. (e) Zeeman splitting of edge states as a function of magnetic field angle with respect to 1D edge ($Y$ axis), where $g\mu_B|B|$ is fixed to be 1.8 meV. Inset shows the angle between magnetic field and 1D edge (thick black line). (f) Phase diagram of the existence regime for MZMs at the corner as function of chemical potential $\mu$ and magnetic field angle $\theta$. The up(down)-pointing triangles represent the TPT line for (100)((010)) edge. In the pink/blue regimes, the helical edge states at two perpendicular edges have different/the same topological characters, thus can/cannot host MZMs at the corner. The superconducting gap $\Delta_1$ is set to be $1.3$ meV and the chemical potential is chosen relative to the Dirac point of edge states. \label{phasediagram} }
\end{figure}

To study the edge properties of a Fe(Se,Te) monolayer, we consider a semi-infinite system for the above Bogoliubov-de Gennes (BdG) type tight-binding Hamiltonian $\mathcal{H}_0+\mathcal{H}_{SC}+\mathcal{H}_Z$ for open boundary conditions. We consider the (100) edge (with the open boundary condition along the X direction in Fig. \ref{setup}(e)) and the angle between the magnetic field and the 1D edge is labelled by $\theta$ in the inset of Fig.\ref{phasediagram}(e). Figs. \ref{phasediagram} (a)-(c) show the edge energy spectrum along $\Gamma-Y$ for different X-directional magnetic fields ($\theta=90^\circ$). Here we adopt the $s_{\pm}$-wave pairing with SC parameters $\Delta_0=0$ and $\Delta_1=1.3$ meV and the magnetic field dependent SC gap is assumed to be $\Delta_0(B)=\Delta_0\sqrt{1-\frac{B^2}{B^2_c}}$\cite{Douglass1961}, where $B_c=45$ T is the upper critical field. We find an SC gap at $\Gamma$ for helical edge states at zero magnetic field (Fig. \ref{phasediagram} (a)). An $s$-wave pairing with non-zero $\Delta_0$ will only have quantitative effects our results as the edge Dirac cone is located around $\Gamma$. Upon increasing the magnetic field to $\mu_Bg_0B_X=1.4$ meV, the Kramers degeneracy at $\Gamma$ is split due to time reversal symmetry breaking. One branch of the bands close the SC gap at $\Gamma$ and form a gapless mode with linear dispersion (Fig. \ref{phasediagram} (b)). The corresponding low energy effective theory is equivalent to the Kitaev model for 1D p-wave spinless SCs (see Sec. III and IV of SM), suggesting a TPT with 1D gapless Majorana mode occurring at this gap closing. Further increasing the magnetic field ($\mu_Bg_0B_X=2.8$ meV) makes the gap re-open. The magnetic gap dominates over the SC gap (Fig. \ref{phasediagram} (c)), thus driving the system into a topologically distinct phase from that in Fig. \ref{phasediagram} (a).

We track the gap evolution as a function of magnetic fields $B$ and chemical potentials $\mu$ in Fig. \ref{phasediagram} (d). The band structures in Figs. \ref{phasediagram} (a)-(c) correspond to the red stars a,b,c in the $B-\mu$ phase diagram of Fig. \ref{phasediagram} (d). A gap closing line separates two topologically distinct phases, one dominated by a magnetic gap and the other dominated by the SC gap, labeled as the phases I and II in Fig. \ref{phasediagram} (d). The existence of the TPT at the edge of a Fe(Se,Te) monolayer suggests that the MZMs can exist at the domain wall between the phases I and II. This scenario was previously discussed for other QSH systems in proximity to SCs\cite{Fu2008,Alicea2012}. We find a minimal value of $B$ field relating to 1.4 meV, i.e., corresponding to a magnetic field of 12 T assuming $g_0=2$, for the TPT line at $\mu=0$ meV. This minimal value is set by the SC gap and thus a large enough magnetic field is required to achieve the phase I. Fortunately, this magnetic field is still well below the in-plane critical magnetic field $B_c\sim 45$ T of a Fe(Se,Te) monolayer \cite{Salamon2016}. 

The magnetic gap further depends on the angle $\theta$ in Fig.\ref{phasediagram}(e). As $\theta$ rotates from 0$^\circ$ to 90$^\circ$, the gap induced solely by Zeeman coupling monotonically increases from 1.1 meV to 2.5 meV in Fig.\ref{phasediagram}(e) for a fixed magnetic field amplitude $|g\mu_BB|=1.8$ meV and $\Delta_1=0$. The anisotropy of the magnetic gap between perpendicular and parallel magnetic fields is significant $\delta_M=\frac{V_{max}-V_{min}}{V_{avg}}\sim 78\%$. As a consequence, the TPT line for the parallel magnetic field ($\theta=0^\circ$) is different from that of the perpendicular magnetic field ($\theta=90^\circ$), as shown by the dashed line in Fig. \ref{phasediagram} (d). We further calculate the TPT lines for the two orthogonal edges (the edges along the X and Y directions) as a function of the chemical potential $\mu$ and the field direction angle $\theta$ with a fixed field amplitude $|g\mu_BB|=1.8$ meV and $\Delta_1=1.3$ meV, as shown in Fig.\ref{phasediagram}(f). Topological properties of two edges are the same (distinct) in the blue (pink) regions. The existence of magnetic anisotropy for the edge states is essential for the MZMs at the corner discussed below.

\textit{ Origin of magnetic anisotropy}
We further study the origin magnetic anisotropy for the edge states in Fe(Te,Se) monolayer. To capture topological property of Fe(Se,Te) monolayer, the odd-parity $j_z=\pm\frac{1}{2}$ and even-parity $j_z=\pm\frac{3}{2}$ states at $\Gamma$ needs to be included in the effective Hamiltonian while the even-parity $j_z=\pm\frac{1}{2}$ state is omitted although it gives the highest valence band \cite{Wu2016} (See the analysis in SM). On the basis functions $\psi^\dag_{\bm{k}}=(c^\dag_{\bm{k}\frac{1}{2}},c^\dag_{\bm{k}\frac{3}{2}},c^\dag_{\bm{k}-\frac{1}{2}},c^\dag_{\bm{k}-\frac{3}{2}})$,
the effective Hamiltonian takes the form of Bernevig-Hughes-Zhang (BHZ) model \cite{Bernevig2006,Wu2016}, given by $\mathcal{H}_{BHZ}=\sum_{\bm{k}}\psi^\dag_{\bm{k}}h_0(\bm{k})\psi_{\bm{k}}$ with 
\begin{eqnarray}
h_0(\bm{k})&=&\epsilon_0(\mathbf{k}) + M(\mathbf{k})\sigma_3+A(k_Y s_0\sigma_1+k_Xs_3\sigma_2),
\label{TI2d}
\end{eqnarray}
where $\epsilon_0(\mathbf{k})=C-D(k^2_X+k^2_Y)$, $M(k)=M-B(k^2_X+k^2_Y)$, $s_l$ and $\sigma_l$ label the Pauli matrices in the pseudo-spin and pseudo-orbital spaces, and $C,D,M,B,A$ are material dependent parameters. The inplane Zeeman coupling $\mathcal{H}_Z$ can also be projected into the basis functions $\psi^\dag_{\bm{k}}$ and is transformed to
$h_Z=\mu_B(g_1P_{1/2}+g_2P_{3/2})B_Xs_1+\mu_B(g_1P_{1/2}-g_2P_{3/2})B_Ys_2 $, where $P_{1/2(3/2)}=(\sigma_0+(-)\sigma_3)/2$ is the projector operator in the subspace of the $j_z=\pm 1/2 (\pm 3/2)$ states and $g_{1,2}$ are effective g-factors for the BHZ model.
To investigate the effective Zeeman coupling of helical edge states, we first calculate the helical edge state from the BHZ Hamiltonian. For the (100) edge, we can omit the $\epsilon_0(\bm{k})$ term and adopt $k_X\rightarrow -i\partial_X$. The Hamiltonian reads
\begin{eqnarray}
h_0(-i\partial_X,k_Y)&=& [M-B(-\partial^2_X+k^2_Y)]\sigma_3\nonumber\\
&+&A(k_Y s_0\sigma_1-i\partial_Xs_3\sigma_2),
\label{TI2d}
\end{eqnarray}
of which the eigenvalue equation $h_0\Psi_p(X)=E_p\Psi_p(X)$ at $k_Y=0$ under the boundary condition $\Psi_p(X\rightarrow 0)=\Psi_p(X\rightarrow +\infty)=0$ can be solved and two zero-energy mode solutions are given by
\begin{eqnarray}
\Psi_p(X)=Nsinh(\eta_1X)e^{\eta_2X}\phi_p,
\end{eqnarray}
where the normalization factor $N=4|\eta_2(\eta_2^2-\eta_1^2)/\eta_1^2|$, $\eta_1=\sqrt{\frac{A^2}{4B^2}-\frac{M}{B}}$ and $\eta_2=\frac{A}{2B}$. The eigenvectors $\phi_p$ satisfy the eigenequation $s_3\sigma_1\phi_p=\phi_p$ and thus can be written as
\begin{eqnarray}
\phi_1&=&|\sigma_1=+1,s_3=+1\rangle,\\
\phi_2&=&|\sigma_1=-1,s_3=-1\rangle.
\end{eqnarray}
The effective Zeeman coupling can be projected into the subspace of helical edge states $\Psi_p(X)$ and the effective Zeeman coupling for helical edge states under an inplane magnetic field is given by
\begin{eqnarray}
h_{edge}(k_Y)&=&\tilde{A}k_Y \tilde{s}_3+\mu_Bg_{E,X}B_X \tilde{s}_1+\mu_Bg_{E,Y}B_Y\tilde{s}_2,
\end{eqnarray}
with $\tilde{s}_i$ is the Pauli matrix in helical edge states space and the effective g-factors for edge states are $g_{E,X}=(g_{1}-g_{2})/2$ and $g_{E,Y}=(g_{1}+g_{2})/2$ (more details can be found SM). From the above form of Zeeman coupling, the non-zero values of both $g_{1}$ and $g_{2}$ make the magnetic gaps of helical edge states different between the parallel and perpendicular magnetic field direction with respect to the edge direction. The anisotropic Zeeman splitting of helical edge states from aforementioned tight binding model calculations can be reproduced in the effective model (see SM). As the orbital Zeeman term can increase $g_{2}$, it will also enhance the anisotropy of Zeeman splitting for edge states.  The inplane Zeeman coupling is isotropic in the bulk but anisotropic on edges, and such a unique magnetic anisotropy is directly derived from its basis with different total angular momenta $j_z=\pm\frac{1}{2},\pm\frac{3}{2}$ of the BHZ model.

\begin{figure}[tb]
\centerline{\includegraphics[width=1\columnwidth]{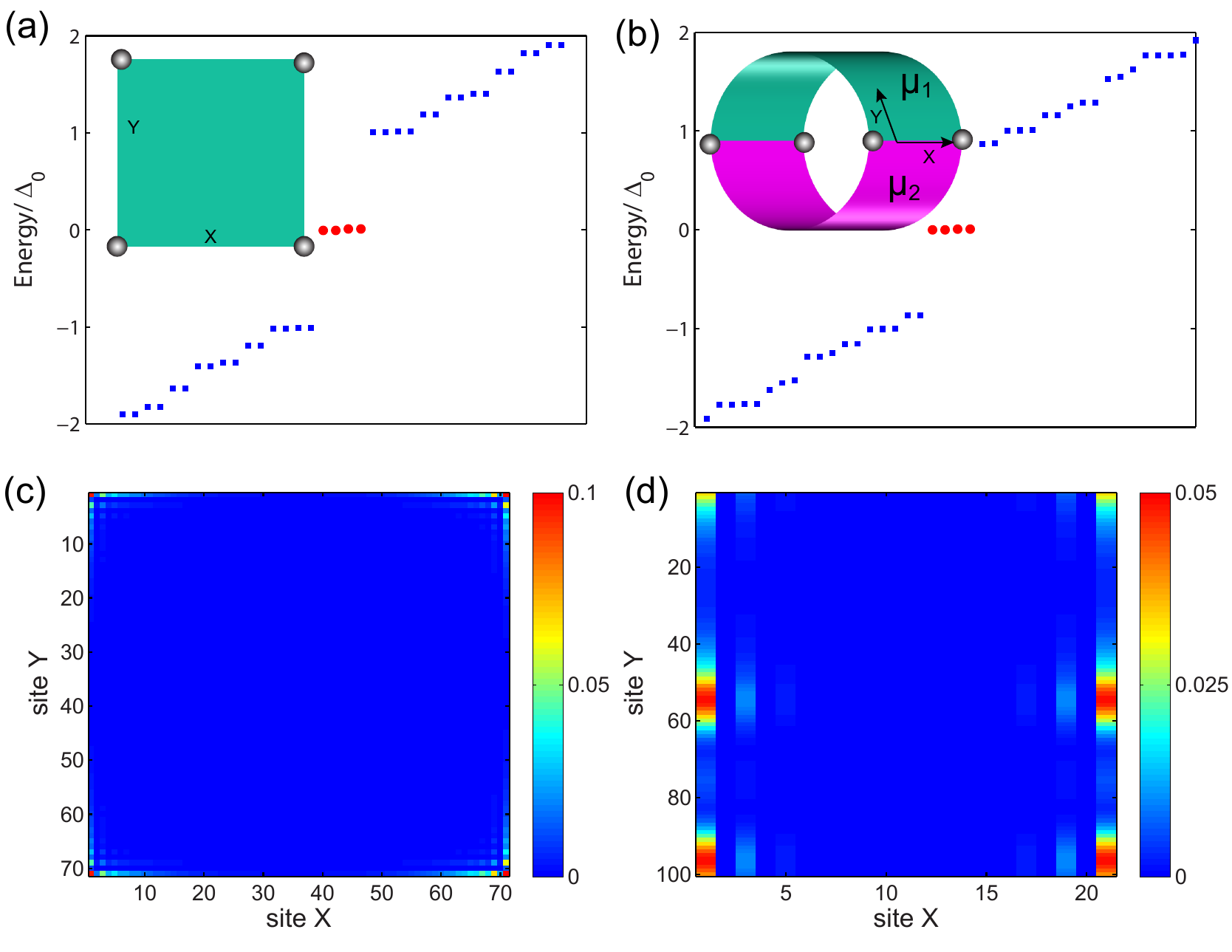}}
\caption{ Energy spectra and spatial pattern for
MZMs on the square and CPDW geometries. (a) The energy spectra and (b) the wavefunctions of MZMs are shown for a square geometry with four MZMs labeled by the gray circles in the inset of (a). (c) The energy spectra and (d) the wavefunctions of MZMs are shown for the CPDW geometry with the periodic boundary condition along $Y$ axis and the open boundary condition along $X$ in the inset of (d), in which the locations of four MZMs are also denoted by the gray circles. We adopted the parameters given Sec. V in the SM.
\label{gapanisotropyMajorana} }
\end{figure}

\textit{ MZMs at the corner and the edge CPDW}
 Due to the existence of a TPT at the edge, MZMs can appear at the domain wall between the phases I and II. To explicitly demonstrate this scenario, we compute its energy levels and show the existence of the MZMs in two different experimental configurations (Fig.\ref{setup}(b) and (c), as well the insets in Fig. \ref{gapanisotropyMajorana}(a) and (b)) based on the effective Hamiltonian $h_0+h_Z$. In the superconducting phase, the BdG Hamiltonian, for the basis $\Psi^\dag_{\bm{k}}=(\psi^\dag_{\bm{k}},\psi^T_{-\bm{k}})$, is given by
\begin{eqnarray}
&&\mathcal{H}_{BdG}=\sum_{\bm{k}}\Psi^\dag_{\bm{k}}h_{BdG}(\mathbf{k})\Psi_{\bm{k}}, \\
&&h_{BdG}= \nonumber\\
&& \left(\begin{array}{cc}
h_{0}(\mathbf{k})+h_Z(-\bm{k})-\mu & {\Delta}(\bm{k}) \\
{\Delta}^\dag(\bm{k}) & -h^*_{0}(-\mathbf{k})-h^*_Z(-\bm{k})+\mu \\
\end{array}\right),\nonumber
\end{eqnarray}
with ${\Delta}(\bm{k})=[\Delta_0+\Delta_1-\Delta_1/4(k^2_X+k^2_Y)]s_2\sigma_3$ for extended $s$-wave ($s_{\pm}$) pairing. We notice that the superconducting gap is opposite for two orbitals in the BHZ model. The detailed derivation of $h_0, h_Z$ and $\Delta$ is provided in the SM. By choosing appropriate parameters $C,D,M,B,A,g_1,g_2$, the effective model can well reproduce the band structure of the tight-binding model near $\Gamma$ and anisotropic Zeeman splitting (see SM). As the topological properties in Fe(Te,Se) monolayers are dominated by the electronic structures around $\Gamma$ point, where $s$- and  $s_{\pm}$-wave pairings exhibit similar behaviors, Majorana states are robust and irrespective of pairing symmetries in the system.

This model allows us to directly calculate the MZMs in two configurations as depicted in the inset of Fig. \ref{gapanisotropyMajorana}(a) and (b). We first consider a square geometry with four corners and apply a magnetic field along the X direction ($\theta = $90$^\circ$ relative to the (100) edge). The parameter choice of the BHZ model is discussed and provided in the Sec. V of the SM, for which we find the (100) edge in phase I while the (010) edge is in phase II.
As a result, four MZMs appear at zero energy and are well localized at the four corners, as shown in Fig.\ref{gapanisotropyMajorana}(a) and (c). The appearance of MZMs at the corner also implies that the bulk SC represents a higher order TSC phase\cite{YanZB2018,WangQY2018,WangYX2018,pan2018lattice,zhu2018tunable,geier2018second,khalaf2018higher,wu2019higher,ZhangRX2019PRL,Volpez2019,Ghorashi2019}, for which our system  provides a concise experimental platform in an intrinsic and high-$T_c$ SC domain. The localization length of the MZM depends on the velocity of edge states and the superconducting gap. By choosing $v_f=4.4\times 10^3 m/s$ and $\Delta_{\text{edge}}=0.5-1$ meV\cite{Zhang2018}, we estimate the localization length as $\hbar v_f/\Delta_{\text{edge}}\sim$ $30-60$\AA~, which can be conveniently measured in experiments.
Similar calculations can also be performed for the CPDW in a slab configuration with the open boundary condition along the X direction and the periodic boundary condition along Y, as shown in the inset of Fig. \ref{gapanisotropyMajorana} (b).
By carefully choosing the chemical potentials $\mu_1$ and $\mu_2$, the green and pink regions in the inset of \ref{gapanisotropyMajorana} (b) are in the phases I and II, respectively, thus allowing for MZM located at two ends of the CPDW between these two regions. Our calculations in Fig. \ref{gapanisotropyMajorana} (b) and (d) indeed show four MZMs appearing at zero energy due to two CPDWs in one period of the whole system.

\begin{figure}[tb]
\centerline{\includegraphics[width=1\columnwidth]{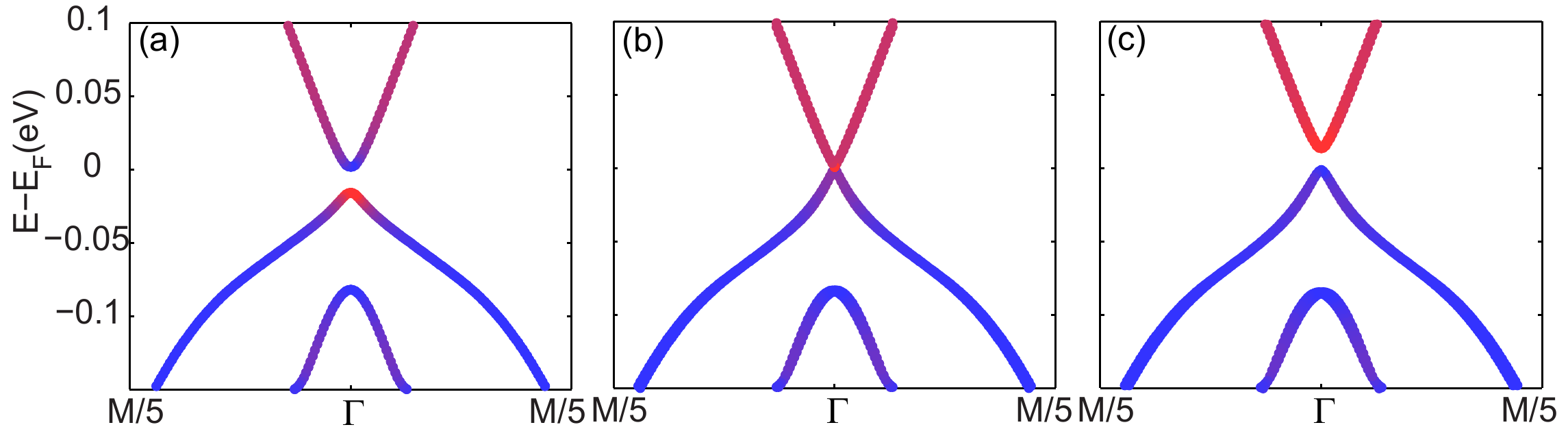}}
\caption{ 2D Bulk TPT for Fe(Te,Se) monolayer as a function of the asymmetric potential induced by the dual gate voltage. The band structure is (a)inverted without electric field, (b) at the critical gapless point at $V_g=0.3$ eV, and (c) normal at $V_g=0.4$ eV. The red color is for the $p_z$ orbitals of Te/Se while the blue for the orbitals from Fe atoms.
 \label{bandinversion} }
\end{figure}

\textit{ Electric field induced 2D bulk TPT}
Finally, an additional direction to achieve MZMs inside the 2D bulk system is given through patterning local gating to form a tri-junction with three different regions, as labelled by $\mu_1$, $\mu_2$ and $V_{g3}$ in Fig.\ref{setup}(d). The situation here is quite similar to the edge CPDW in Fig.\ref{setup}(c) and Fig. \ref{gapanisotropyMajorana}(b). We choose the same chemical potentials $\mu_1$ and $\mu_2$ for two regions of the tri-junction to be in the phases I and II in Fig. \ref{phasediagram} (d). If we can achieve a trivial insulator phase in the last region $V_{g3}$ of Fig.\ref{setup}(d), which is equivalent to the vacuum termination in Fig.\ref{setup}(c), the tri-junction is topologically equivalent to the edge CPDW and thus allows for the existence of MZMs. Therefore, we next show the existence of a 2D bulk TPT induced by electric gating to tune the 2D Fe(Se,Te) monolayer between a QSH phase and a trivial insulator phase. The key idea here is that the $p_z$ orbital of Se or Te atoms is strongly hybridized with the Fe $d_{xy}$ orbital, and thus contributes significantly to the odd-parity $j_z=\pm\frac{1}{2}$ bands, but not to the even-parity $j_z=\pm\frac{3}{2}$ bands\cite{Wu2016}. Since the Fe layer is sandwiched between two Se (or Te) layers in an Fe(Se,Te) monolayer, the asymmetric potential between two Se (or Te) layers can induce an energy shift of the $j_z=\pm\frac{1}{2}$ bands with respect to the $j_z=\pm\frac{3}{2}$ bands. Thus, if we initially tune the band gap close to zero by controlling the Se/Te composition ratio, the 2D bulk TPT between the QSH state and the trivial insulator can be induced by a dual gate voltage. Our tight-binding model does not explicitly involve $p_z$ orbital of Se or Te atoms and thus not ideally suitable for studying this mechanism. Instead, we perform a calculation based on the tight-binding model including both Fe $d$ and Se/Te $p$ oribtals from the Wannier function method\cite{Mostofi2014} (See SM for more technical details). The energy dispersions are shown in Fig. \ref{bandinversion}(a)-(c) for different asymmetric potentials. The QSH state in Fig. \ref{bandinversion}(a) and the trivial insulator phase in Fig. \ref{bandinversion}(c) are separated by a 2D TPT shown in Fig. \ref{bandinversion}(b). The band inversion can be further revealed by projecting each band to the atomic orbitals. The red and blue colors in Fig. \ref{bandinversion}(a)-(c) represent the atomic orbital contribution from Fe and Se/Te atoms, from which one can easily see the inverted band structure in Fig. \ref{bandinversion}(a) and a trivial insulator phase in Fig. \ref{bandinversion}(c). This band inversion is induced by the inversion symmetry breaking due to the electric gates, as discussed in SM. In summary, by making the experimental setup shown in Fig.\ref{setup}(d), our study indicated that MZMs can indeed be realized in a tri-junction configuration not just limited to corners.

{\textit{Conclusion:}}
Our work theoretically demonstrates the suitability of a high-T$_c$ SC Fe(Te,Se) monolayer as a platform for the realization of MZMs. This is the first theoretical proposal about the intrinsic material realization of time-reversal-breaking high-order topological superconductivity with Majorana corner modes in iron-based superconductor systems. Moreover, the scenario of edge magnetic anisotropy, even surviving with an $s$-wave pairing, serves as a general principle to search for high-order topological superconductivity in other materials. A magnetic layer, such as CrX$_3$ (X=I,Br,Cl)\cite{Huang2018,Ghazaryan2018,Abramchuk2018,McGuire2017}, CrGeTe$_3$\cite{Gong2017} or FeTe\cite{dai2015antiferromagnetic}, can be inserted between the Fe(Te,Se) monolayer and the insulator layers in Fig. \ref{setup} (a) in order to enhance the magnetic gap of helical edge states through magnetic proximity effect. While the underlying mechanism is similar, this may broaden the parameter regimes for MZMs because of the much stronger exchange interaction compared to the Zeeman coupling. In addition, Josephson junctions may provide an alternative approach to realize MZMs in a Fe(Te,Se) monolayer \cite{Pientka2017}, an approach which has recently been applied to semiconductor/SC heterostructures \cite{RenHC2019,FornieriA2019}. Furthermore, rotating the inplane
magnetic field may provide an efficient approach to perform the braiding operation for the corner MZMs\cite{Pahomi2019}. The 2D nature also makes our platform suitable for the potential manipulation and detection of MZMs the implementation scheme of which we leave for future work.


{\it Acknowledgement: }
C.X.Liu acknowledges the support of the Office of Naval Research (Grant No. N00014-18-1-2793), DOE grant (DE-SC0019064) and Kaufman New Initiative research grant of the Pittsburgh Foundation. The work in W\"urzburg is supported by the German Research Foundation (DFG) through DFG-SFB 1170, project B04 and by the W\"urzburg-Dresden Cluster of Excellence on Complexity and Topology in Quantum Matter -- \textit{ct.qmat} (EXC 2147, project-id 39085490).

{\it Note added: } We recently became aware of two independent works on Majorana corner modes, one by Zhang {\it el at} \cite{ZhangRX2019} for the Fe(Se,Te)/FeTe heterostructure and the other by Wu {\it et al} \cite{wu2019inplane} for the QSH/SC heterostructure under in-plane magnetic fields.

%
%


\bibliography{FeTeSe_TSCrefNC}

\begin{thebibliography}{69}%
\makeatletter
\providecommand \@ifxundefined [1]{%
 \@ifx{#1\undefined}
}%
\providecommand \@ifnum [1]{%
 \ifnum #1\expandafter \@firstoftwo
 \else \expandafter \@secondoftwo
 \fi
}%
\providecommand \@ifx [1]{%
 \ifx #1\expandafter \@firstoftwo
 \else \expandafter \@secondoftwo
 \fi
}%
\providecommand \natexlab [1]{#1}%
\providecommand \enquote  [1]{``#1''}%
\providecommand \bibnamefont  [1]{#1}%
\providecommand \bibfnamefont [1]{#1}%
\providecommand \citenamefont [1]{#1}%
\providecommand \href@noop [0]{\@secondoftwo}%
\providecommand \href [0]{\begingroup \@sanitize@url \@href}%
\providecommand \@href[1]{\@@startlink{#1}\@@href}%
\providecommand \@@href[1]{\endgroup#1\@@endlink}%
\providecommand \@sanitize@url [0]{\catcode `\\12\catcode `\$12\catcode
  `\&12\catcode `\#12\catcode `\^12\catcode `\_12\catcode `\%12\relax}%
\providecommand \@@startlink[1]{}%
\providecommand \@@endlink[0]{}%
\providecommand \url  [0]{\begingroup\@sanitize@url \@url }%
\providecommand \@url [1]{\endgroup\@href {#1}{\urlprefix }}%
\providecommand \urlprefix  [0]{URL }%
\providecommand \Eprint [0]{\href }%
\providecommand \doibase [0]{http://dx.doi.org/}%
\providecommand \selectlanguage [0]{\@gobble}%
\providecommand \bibinfo  [0]{\@secondoftwo}%
\providecommand \bibfield  [0]{\@secondoftwo}%
\providecommand \translation [1]{[#1]}%
\providecommand \BibitemOpen [0]{}%
\providecommand \bibitemStop [0]{}%
\providecommand \bibitemNoStop [0]{.\EOS\space}%
\providecommand \EOS [0]{\spacefactor3000\relax}%
\providecommand \BibitemShut  [1]{\csname bibitem#1\endcsname}%
\let\auto@bib@innerbib\@empty
\bibitem [{\citenamefont {Peng}\ \emph {et~al.}(2019)\citenamefont {Peng},
  \citenamefont {Li}, \citenamefont {Wu}, \citenamefont {Deng}, \citenamefont
  {Shi}, \citenamefont {Fan}, \citenamefont {Li}, \citenamefont {Huang},
  \citenamefont {Qian}, \citenamefont {Richard}, \citenamefont {Hu},
  \citenamefont {Pan}, \citenamefont {Mao}, \citenamefont {Sun},\ and\
  \citenamefont {Ding}}]{Peng2019}%
  \BibitemOpen
  \bibfield  {author} {\bibinfo {author} {\bibfnamefont {X.-L.}\ \bibnamefont
  {Peng}}, \bibinfo {author} {\bibfnamefont {Y.}~\bibnamefont {Li}}, \bibinfo
  {author} {\bibfnamefont {X.-X.}\ \bibnamefont {Wu}}, \bibinfo {author}
  {\bibfnamefont {H.-B.}\ \bibnamefont {Deng}}, \bibinfo {author}
  {\bibfnamefont {X.}~\bibnamefont {Shi}}, \bibinfo {author} {\bibfnamefont
  {W.-H.}\ \bibnamefont {Fan}}, \bibinfo {author} {\bibfnamefont
  {M.}~\bibnamefont {Li}}, \bibinfo {author} {\bibfnamefont {Y.-B.}\
  \bibnamefont {Huang}}, \bibinfo {author} {\bibfnamefont {T.}~\bibnamefont
  {Qian}}, \bibinfo {author} {\bibfnamefont {P.}~\bibnamefont {Richard}},
  \bibinfo {author} {\bibfnamefont {J.-P.}\ \bibnamefont {Hu}}, \bibinfo
  {author} {\bibfnamefont {S.-H.}\ \bibnamefont {Pan}}, \bibinfo {author}
  {\bibfnamefont {H.-Q.}\ \bibnamefont {Mao}}, \bibinfo {author} {\bibfnamefont
  {Y.-J.}\ \bibnamefont {Sun}}, \ and\ \bibinfo {author} {\bibfnamefont
  {H.}~\bibnamefont {Ding}},\ }\href@noop {} {\bibfield  {journal} {\bibinfo
  {journal} {Phys. Rev. B}\ }\textbf {\bibinfo {volume} {100}},\ \bibinfo
  {pages} {155134} (\bibinfo {year} {2019})}\BibitemShut {NoStop}%
\bibitem [{\citenamefont {Ivanov}(2001)}]{Ivanov2001}%
  \BibitemOpen
  \bibfield  {author} {\bibinfo {author} {\bibfnamefont {D.~A.}\ \bibnamefont
  {Ivanov}},\ }\href@noop {} {\bibfield  {journal} {\bibinfo  {journal} {Phys.
  Rev. Lett.}\ }\textbf {\bibinfo {volume} {86}},\ \bibinfo {pages} {268}
  (\bibinfo {year} {2001})}\BibitemShut {NoStop}%
\bibitem [{\citenamefont {Kitaev}(2003)}]{Kitaev2003}%
  \BibitemOpen
  \bibfield  {author} {\bibinfo {author} {\bibfnamefont {A.~Y.}\ \bibnamefont
  {Kitaev}},\ }\href@noop {} {\bibfield  {journal} {\bibinfo  {journal} {Ann.
  Phys. (N. Y.)}\ }\textbf {\bibinfo {volume} {303}},\ \bibinfo {pages} {2}
  (\bibinfo {year} {2003})}\BibitemShut {NoStop}%
\bibitem [{\citenamefont {Kitaev}(2006)}]{Kitaev2006}%
  \BibitemOpen
  \bibfield  {author} {\bibinfo {author} {\bibfnamefont {A.}~\bibnamefont
  {Kitaev}},\ }\href@noop {} {\bibfield  {journal} {\bibinfo  {journal} {Ann.
  Phys.(N. Y.)}\ }\textbf {\bibinfo {volume} {321}},\ \bibinfo {pages} {2}
  (\bibinfo {year} {2006})}\BibitemShut {NoStop}%
\bibitem [{\citenamefont {Nayak}\ \emph {et~al.}(2008)\citenamefont {Nayak},
  \citenamefont {Simon}, \citenamefont {Stern}, \citenamefont {Freedman},\ and\
  \citenamefont {Das~Sarma}}]{Nayak2008}%
  \BibitemOpen
  \bibfield  {author} {\bibinfo {author} {\bibfnamefont {C.}~\bibnamefont
  {Nayak}}, \bibinfo {author} {\bibfnamefont {S.~H.}\ \bibnamefont {Simon}},
  \bibinfo {author} {\bibfnamefont {A.}~\bibnamefont {Stern}}, \bibinfo
  {author} {\bibfnamefont {M.}~\bibnamefont {Freedman}}, \ and\ \bibinfo
  {author} {\bibfnamefont {S.}~\bibnamefont {Das~Sarma}},\ }\href@noop {}
  {\bibfield  {journal} {\bibinfo  {journal} {Rev. Mod. Phys.}\ }\textbf
  {\bibinfo {volume} {80}},\ \bibinfo {pages} {1083} (\bibinfo {year}
  {2008})}\BibitemShut {NoStop}%
\bibitem [{\citenamefont {Alicea}(2012)}]{Alicea2012}%
  \BibitemOpen
  \bibfield  {author} {\bibinfo {author} {\bibfnamefont {J.}~\bibnamefont
  {Alicea}},\ }\href@noop {} {\bibfield  {journal} {\bibinfo  {journal} {Rep.
  Prog. Phys.}\ }\textbf {\bibinfo {volume} {75}},\ \bibinfo {pages} {076501}
  (\bibinfo {year} {2012})}\BibitemShut {NoStop}%
\bibitem [{\citenamefont {Sarma}\ \emph {et~al.}(2015)\citenamefont {Sarma},
  \citenamefont {Freedman},\ and\ \citenamefont {Nayak}}]{Sarma2015}%
  \BibitemOpen
  \bibfield  {author} {\bibinfo {author} {\bibfnamefont {S.~D.}\ \bibnamefont
  {Sarma}}, \bibinfo {author} {\bibfnamefont {M.}~\bibnamefont {Freedman}}, \
  and\ \bibinfo {author} {\bibfnamefont {C.}~\bibnamefont {Nayak}},\
  }\href@noop {} {\bibfield  {journal} {\bibinfo  {journal} {Npj Quantum
  Information}\ }\textbf {\bibinfo {volume} {1}},\ \bibinfo {pages} {15001}
  (\bibinfo {year} {2015})}\BibitemShut {NoStop}%
\bibitem [{\citenamefont {Aasen}\ \emph {et~al.}(2016)\citenamefont {Aasen},
  \citenamefont {Hell}, \citenamefont {Mishmash}, \citenamefont {Higginbotham},
  \citenamefont {Danon}, \citenamefont {Leijnse}, \citenamefont {Jespersen},
  \citenamefont {Folk}, \citenamefont {Marcus}, \citenamefont {Flensberg},\
  and\ \citenamefont {Alicea}}]{Aasen2016}%
  \BibitemOpen
  \bibfield  {author} {\bibinfo {author} {\bibfnamefont {D.}~\bibnamefont
  {Aasen}}, \bibinfo {author} {\bibfnamefont {M.}~\bibnamefont {Hell}},
  \bibinfo {author} {\bibfnamefont {R.~V.}\ \bibnamefont {Mishmash}}, \bibinfo
  {author} {\bibfnamefont {A.}~\bibnamefont {Higginbotham}}, \bibinfo {author}
  {\bibfnamefont {J.}~\bibnamefont {Danon}}, \bibinfo {author} {\bibfnamefont
  {M.}~\bibnamefont {Leijnse}}, \bibinfo {author} {\bibfnamefont {T.~S.}\
  \bibnamefont {Jespersen}}, \bibinfo {author} {\bibfnamefont {J.~A.}\
  \bibnamefont {Folk}}, \bibinfo {author} {\bibfnamefont {C.~M.}\ \bibnamefont
  {Marcus}}, \bibinfo {author} {\bibfnamefont {K.}~\bibnamefont {Flensberg}}, \
  and\ \bibinfo {author} {\bibfnamefont {J.}~\bibnamefont {Alicea}},\
  }\href@noop {} {\bibfield  {journal} {\bibinfo  {journal} {Phys. Rev. X}\
  }\textbf {\bibinfo {volume} {6}},\ \bibinfo {pages} {031016} (\bibinfo {year}
  {2016})}\BibitemShut {NoStop}%
\bibitem [{\citenamefont {Karzig}\ \emph {et~al.}(2017)\citenamefont {Karzig},
  \citenamefont {Knapp}, \citenamefont {Lutchyn}, \citenamefont {Bonderson},
  \citenamefont {Hastings}, \citenamefont {Nayak}, \citenamefont {Alicea},
  \citenamefont {Flensberg}, \citenamefont {Plugge}, \citenamefont {Oreg},
  \citenamefont {Marcus},\ and\ \citenamefont {Freedman}}]{Karzig2017}%
  \BibitemOpen
  \bibfield  {author} {\bibinfo {author} {\bibfnamefont {T.}~\bibnamefont
  {Karzig}}, \bibinfo {author} {\bibfnamefont {C.}~\bibnamefont {Knapp}},
  \bibinfo {author} {\bibfnamefont {R.~M.}\ \bibnamefont {Lutchyn}}, \bibinfo
  {author} {\bibfnamefont {P.}~\bibnamefont {Bonderson}}, \bibinfo {author}
  {\bibfnamefont {M.~B.}\ \bibnamefont {Hastings}}, \bibinfo {author}
  {\bibfnamefont {C.}~\bibnamefont {Nayak}}, \bibinfo {author} {\bibfnamefont
  {J.}~\bibnamefont {Alicea}}, \bibinfo {author} {\bibfnamefont
  {K.}~\bibnamefont {Flensberg}}, \bibinfo {author} {\bibfnamefont
  {S.}~\bibnamefont {Plugge}}, \bibinfo {author} {\bibfnamefont
  {Y.}~\bibnamefont {Oreg}}, \bibinfo {author} {\bibfnamefont {C.~M.}\
  \bibnamefont {Marcus}}, \ and\ \bibinfo {author} {\bibfnamefont {M.~H.}\
  \bibnamefont {Freedman}},\ }\href@noop {} {\bibfield  {journal} {\bibinfo
  {journal} {Phys. Rev. B}\ }\textbf {\bibinfo {volume} {95}},\ \bibinfo
  {pages} {235305} (\bibinfo {year} {2017})}\BibitemShut {NoStop}%
\bibitem [{\citenamefont {Moore}\ and\ \citenamefont {Read}(1991)}]{Moore1991}%
  \BibitemOpen
  \bibfield  {author} {\bibinfo {author} {\bibfnamefont {G.}~\bibnamefont
  {Moore}}\ and\ \bibinfo {author} {\bibfnamefont {N.}~\bibnamefont {Read}},\
  }\href@noop {} {\bibfield  {journal} {\bibinfo  {journal} {Nucl. Phys. B}\
  }\textbf {\bibinfo {volume} {360}},\ \bibinfo {pages} {362} (\bibinfo {year}
  {1991})}\BibitemShut {NoStop}%
\bibitem [{\citenamefont {Read}\ and\ \citenamefont {Green}(2000)}]{Read2000}%
  \BibitemOpen
  \bibfield  {author} {\bibinfo {author} {\bibfnamefont {N.}~\bibnamefont
  {Read}}\ and\ \bibinfo {author} {\bibfnamefont {D.}~\bibnamefont {Green}},\
  }\href@noop {} {\bibfield  {journal} {\bibinfo  {journal} {Phys. Rev. B}\
  }\textbf {\bibinfo {volume} {61}},\ \bibinfo {pages} {10267} (\bibinfo {year}
  {2000})}\BibitemShut {NoStop}%
\bibitem [{\citenamefont {Rice}\ and\ \citenamefont
  {Sigrist}(1995)}]{RiceTM1995}%
  \BibitemOpen
  \bibfield  {author} {\bibinfo {author} {\bibfnamefont {T.~M.}\ \bibnamefont
  {Rice}}\ and\ \bibinfo {author} {\bibfnamefont {M.}~\bibnamefont {Sigrist}},\
  }\href@noop {} {\bibfield  {journal} {\bibinfo  {journal} {J. Phys. Condens.
  Matter.}\ }\textbf {\bibinfo {volume} {7}},\ \bibinfo {pages} {L643}
  (\bibinfo {year} {1995})}\BibitemShut {NoStop}%
\bibitem [{\citenamefont {Das~Sarma}\ \emph {et~al.}(2006)\citenamefont
  {Das~Sarma}, \citenamefont {Nayak},\ and\ \citenamefont
  {Tewari}}]{DasSarma2006}%
  \BibitemOpen
  \bibfield  {author} {\bibinfo {author} {\bibfnamefont {S.}~\bibnamefont
  {Das~Sarma}}, \bibinfo {author} {\bibfnamefont {C.}~\bibnamefont {Nayak}}, \
  and\ \bibinfo {author} {\bibfnamefont {S.}~\bibnamefont {Tewari}},\
  }\href@noop {} {\bibfield  {journal} {\bibinfo  {journal} {Phys. Rev. B}\
  }\textbf {\bibinfo {volume} {73}},\ \bibinfo {pages} {220502} (\bibinfo
  {year} {2006})}\BibitemShut {NoStop}%
\bibitem [{\citenamefont {Greiter}\ and\ \citenamefont
  {Thomale}(2009)}]{PhysRevLett.102.207203}%
  \BibitemOpen
  \bibfield  {author} {\bibinfo {author} {\bibfnamefont {M.}~\bibnamefont
  {Greiter}}\ and\ \bibinfo {author} {\bibfnamefont {R.}~\bibnamefont
  {Thomale}},\ }\href@noop {} {\bibfield  {journal} {\bibinfo  {journal} {Phys.
  Rev. Lett.}\ }\textbf {\bibinfo {volume} {102}},\ \bibinfo {pages} {207203}
  (\bibinfo {year} {2009})}\BibitemShut {NoStop}%
\bibitem [{\citenamefont {Kitaev}(2001)}]{Kitaev2001PU}%
  \BibitemOpen
  \bibfield  {author} {\bibinfo {author} {\bibfnamefont {A.~Y.}\ \bibnamefont
  {Kitaev}},\ }\href@noop {} {\bibfield  {journal} {\bibinfo  {journal}
  {Physics-Uspekhi}\ }\textbf {\bibinfo {volume} {44}},\ \bibinfo {pages} {131}
  (\bibinfo {year} {2001})}\BibitemShut {NoStop}%
\bibitem [{\citenamefont {Lutchyn}\ \emph {et~al.}(2010)\citenamefont
  {Lutchyn}, \citenamefont {Sau},\ and\ \citenamefont
  {Das~Sarma}}]{Lutchyn2010}%
  \BibitemOpen
  \bibfield  {author} {\bibinfo {author} {\bibfnamefont {R.~M.}\ \bibnamefont
  {Lutchyn}}, \bibinfo {author} {\bibfnamefont {J.~D.}\ \bibnamefont {Sau}}, \
  and\ \bibinfo {author} {\bibfnamefont {S.}~\bibnamefont {Das~Sarma}},\
  }\href@noop {} {\bibfield  {journal} {\bibinfo  {journal} {Phys. Rev. Lett.}\
  }\textbf {\bibinfo {volume} {105}},\ \bibinfo {pages} {077001} (\bibinfo
  {year} {2010})}\BibitemShut {NoStop}%
\bibitem [{\citenamefont {Oreg}\ \emph {et~al.}(2010)\citenamefont {Oreg},
  \citenamefont {Refael},\ and\ \citenamefont {von
  Oppen}}]{PhysRevLett.105.177002}%
  \BibitemOpen
  \bibfield  {author} {\bibinfo {author} {\bibfnamefont {Y.}~\bibnamefont
  {Oreg}}, \bibinfo {author} {\bibfnamefont {G.}~\bibnamefont {Refael}}, \ and\
  \bibinfo {author} {\bibfnamefont {F.}~\bibnamefont {von Oppen}},\ }\href@noop
  {} {\bibfield  {journal} {\bibinfo  {journal} {Phys. Rev. Lett.}\ }\textbf
  {\bibinfo {volume} {105}},\ \bibinfo {pages} {177002} (\bibinfo {year}
  {2010})}\BibitemShut {NoStop}%
\bibitem [{\citenamefont {Sau}\ \emph {et~al.}(2010)\citenamefont {Sau},
  \citenamefont {Lutchyn}, \citenamefont {Tewari},\ and\ \citenamefont
  {Das~Sarma}}]{Sau2010}%
  \BibitemOpen
  \bibfield  {author} {\bibinfo {author} {\bibfnamefont {J.~D.}\ \bibnamefont
  {Sau}}, \bibinfo {author} {\bibfnamefont {R.~M.}\ \bibnamefont {Lutchyn}},
  \bibinfo {author} {\bibfnamefont {S.}~\bibnamefont {Tewari}}, \ and\ \bibinfo
  {author} {\bibfnamefont {S.}~\bibnamefont {Das~Sarma}},\ }\href@noop {}
  {\bibfield  {journal} {\bibinfo  {journal} {Phys. Rev. Lett.}\ }\textbf
  {\bibinfo {volume} {104}},\ \bibinfo {pages} {040502} (\bibinfo {year}
  {2010})}\BibitemShut {NoStop}%
\bibitem [{\citenamefont {Alicea}(2010)}]{AliceaJ2010}%
  \BibitemOpen
  \bibfield  {author} {\bibinfo {author} {\bibfnamefont {J.}~\bibnamefont
  {Alicea}},\ }\href@noop {} {\bibfield  {journal} {\bibinfo  {journal} {Phys.
  Rev. B}\ }\textbf {\bibinfo {volume} {81}},\ \bibinfo {pages} {125318}
  (\bibinfo {year} {2010})}\BibitemShut {NoStop}%
\bibitem [{\citenamefont {Fu}\ and\ \citenamefont {Kane}(2008)}]{Fu2008}%
  \BibitemOpen
  \bibfield  {author} {\bibinfo {author} {\bibfnamefont {L.}~\bibnamefont
  {Fu}}\ and\ \bibinfo {author} {\bibfnamefont {C.~L.}\ \bibnamefont {Kane}},\
  }\href@noop {} {\bibfield  {journal} {\bibinfo  {journal} {Phys. Rev. Lett.}\
  }\textbf {\bibinfo {volume} {100}},\ \bibinfo {pages} {096407} (\bibinfo
  {year} {2008})}\BibitemShut {NoStop}%
\bibitem [{\citenamefont {Qi}\ \emph {et~al.}(2010)\citenamefont {Qi},
  \citenamefont {Hughes},\ and\ \citenamefont {Zhang}}]{Qi2010}%
  \BibitemOpen
  \bibfield  {author} {\bibinfo {author} {\bibfnamefont {X.-L.}\ \bibnamefont
  {Qi}}, \bibinfo {author} {\bibfnamefont {T.~L.}\ \bibnamefont {Hughes}}, \
  and\ \bibinfo {author} {\bibfnamefont {S.-C.}\ \bibnamefont {Zhang}},\
  }\href@noop {} {\bibfield  {journal} {\bibinfo  {journal} {Phys. Rev. B}\
  }\textbf {\bibinfo {volume} {82}},\ \bibinfo {pages} {184516} (\bibinfo
  {year} {2010})}\BibitemShut {NoStop}%
\bibitem [{\citenamefont {Chung}\ \emph {et~al.}(2011)\citenamefont {Chung},
  \citenamefont {Qi}, \citenamefont {Maciejko},\ and\ \citenamefont
  {Zhang}}]{Chung2011}%
  \BibitemOpen
  \bibfield  {author} {\bibinfo {author} {\bibfnamefont {S.~B.}\ \bibnamefont
  {Chung}}, \bibinfo {author} {\bibfnamefont {X.-L.}\ \bibnamefont {Qi}},
  \bibinfo {author} {\bibfnamefont {J.}~\bibnamefont {Maciejko}}, \ and\
  \bibinfo {author} {\bibfnamefont {S.-C.}\ \bibnamefont {Zhang}},\ }\href@noop
  {} {\bibfield  {journal} {\bibinfo  {journal} {Phys. Rev. B}\ }\textbf
  {\bibinfo {volume} {83}},\ \bibinfo {pages} {100512} (\bibinfo {year}
  {2011})}\BibitemShut {NoStop}%
\bibitem [{\citenamefont {Nadj-Perge}\ \emph {et~al.}(2013)\citenamefont
  {Nadj-Perge}, \citenamefont {Drozdov}, \citenamefont {Bernevig},\ and\
  \citenamefont {Yazdani}}]{Nadj-Perge2013PRB}%
  \BibitemOpen
  \bibfield  {author} {\bibinfo {author} {\bibfnamefont {S.}~\bibnamefont
  {Nadj-Perge}}, \bibinfo {author} {\bibfnamefont {I.~K.}\ \bibnamefont
  {Drozdov}}, \bibinfo {author} {\bibfnamefont {B.~A.}\ \bibnamefont
  {Bernevig}}, \ and\ \bibinfo {author} {\bibfnamefont {A.}~\bibnamefont
  {Yazdani}},\ }\href@noop {} {\bibfield  {journal} {\bibinfo  {journal} {Phys.
  Rev. B}\ }\textbf {\bibinfo {volume} {88}},\ \bibinfo {pages} {020407}
  (\bibinfo {year} {2013})}\BibitemShut {NoStop}%
\bibitem [{\citenamefont {Braunecker}\ and\ \citenamefont
  {Simon}(2013)}]{Braunecker2013}%
  \BibitemOpen
  \bibfield  {author} {\bibinfo {author} {\bibfnamefont {B.}~\bibnamefont
  {Braunecker}}\ and\ \bibinfo {author} {\bibfnamefont {P.}~\bibnamefont
  {Simon}},\ }\href@noop {} {\bibfield  {journal} {\bibinfo  {journal} {Phys.
  Rev. Lett.}\ }\textbf {\bibinfo {volume} {111}},\ \bibinfo {pages} {147202}
  (\bibinfo {year} {2013})}\BibitemShut {NoStop}%
\bibitem [{\citenamefont {Mourik}\ \emph {et~al.}(2012)\citenamefont {Mourik},
  \citenamefont {Zuo}, \citenamefont {Frolov}, \citenamefont {Plissard},
  \citenamefont {Bakkers},\ and\ \citenamefont {Kouwenhoven}}]{Mourik2012}%
  \BibitemOpen
  \bibfield  {author} {\bibinfo {author} {\bibfnamefont {V.}~\bibnamefont
  {Mourik}}, \bibinfo {author} {\bibfnamefont {K.}~\bibnamefont {Zuo}},
  \bibinfo {author} {\bibfnamefont {S.~M.}\ \bibnamefont {Frolov}}, \bibinfo
  {author} {\bibfnamefont {S.~R.}\ \bibnamefont {Plissard}}, \bibinfo {author}
  {\bibfnamefont {E.~P. A.~M.}\ \bibnamefont {Bakkers}}, \ and\ \bibinfo
  {author} {\bibfnamefont {L.~P.}\ \bibnamefont {Kouwenhoven}},\ }\href@noop {}
  {\bibfield  {journal} {\bibinfo  {journal} {Science}\ }\textbf {\bibinfo
  {volume} {336}},\ \bibinfo {pages} {1003} (\bibinfo {year}
  {2012})}\BibitemShut {NoStop}%
\bibitem [{\citenamefont {Wang}\ \emph {et~al.}(2012)\citenamefont {Wang},
  \citenamefont {Liu}, \citenamefont {Xu}, \citenamefont {Yang}, \citenamefont
  {Miao}, \citenamefont {Yao}, \citenamefont {Gao}, \citenamefont {Shen},
  \citenamefont {Ma}, \citenamefont {Chen}, \citenamefont {Xu}, \citenamefont
  {Liu}, \citenamefont {Zhang}, \citenamefont {Qian}, \citenamefont {Jia},\
  and\ \citenamefont {Xue}}]{WangMX2012}%
  \BibitemOpen
  \bibfield  {author} {\bibinfo {author} {\bibfnamefont {M.-X.}\ \bibnamefont
  {Wang}}, \bibinfo {author} {\bibfnamefont {C.}~\bibnamefont {Liu}}, \bibinfo
  {author} {\bibfnamefont {J.-P.}\ \bibnamefont {Xu}}, \bibinfo {author}
  {\bibfnamefont {F.}~\bibnamefont {Yang}}, \bibinfo {author} {\bibfnamefont
  {L.}~\bibnamefont {Miao}}, \bibinfo {author} {\bibfnamefont {M.-Y.}\
  \bibnamefont {Yao}}, \bibinfo {author} {\bibfnamefont {C.~L.}\ \bibnamefont
  {Gao}}, \bibinfo {author} {\bibfnamefont {C.}~\bibnamefont {Shen}}, \bibinfo
  {author} {\bibfnamefont {X.}~\bibnamefont {Ma}}, \bibinfo {author}
  {\bibfnamefont {X.}~\bibnamefont {Chen}}, \bibinfo {author} {\bibfnamefont
  {Z.-A.}\ \bibnamefont {Xu}}, \bibinfo {author} {\bibfnamefont
  {Y.}~\bibnamefont {Liu}}, \bibinfo {author} {\bibfnamefont {S.-C.}\
  \bibnamefont {Zhang}}, \bibinfo {author} {\bibfnamefont {D.}~\bibnamefont
  {Qian}}, \bibinfo {author} {\bibfnamefont {J.-F.}\ \bibnamefont {Jia}}, \
  and\ \bibinfo {author} {\bibfnamefont {Q.-K.}\ \bibnamefont {Xue}},\
  }\href@noop {} {\bibfield  {journal} {\bibinfo  {journal} {Science}\ }\textbf
  {\bibinfo {volume} {336}},\ \bibinfo {pages} {52} (\bibinfo {year}
  {2012})}\BibitemShut {NoStop}%
\bibitem [{\citenamefont {Nadj-Perge}\ \emph {et~al.}(2014)\citenamefont
  {Nadj-Perge}, \citenamefont {Drozdov}, \citenamefont {Li}, \citenamefont
  {Chen}, \citenamefont {Jeon}, \citenamefont {Seo}, \citenamefont {MacDonald},
  \citenamefont {Bernevig},\ and\ \citenamefont {Yazdani}}]{Nadj-Perge2014}%
  \BibitemOpen
  \bibfield  {author} {\bibinfo {author} {\bibfnamefont {S.}~\bibnamefont
  {Nadj-Perge}}, \bibinfo {author} {\bibfnamefont {I.~K.}\ \bibnamefont
  {Drozdov}}, \bibinfo {author} {\bibfnamefont {J.}~\bibnamefont {Li}},
  \bibinfo {author} {\bibfnamefont {H.}~\bibnamefont {Chen}}, \bibinfo {author}
  {\bibfnamefont {S.}~\bibnamefont {Jeon}}, \bibinfo {author} {\bibfnamefont
  {J.}~\bibnamefont {Seo}}, \bibinfo {author} {\bibfnamefont {A.~H.}\
  \bibnamefont {MacDonald}}, \bibinfo {author} {\bibfnamefont {B.~A.}\
  \bibnamefont {Bernevig}}, \ and\ \bibinfo {author} {\bibfnamefont
  {A.}~\bibnamefont {Yazdani}},\ }\href@noop {} {\bibfield  {journal} {\bibinfo
   {journal} {Science}\ }\textbf {\bibinfo {volume} {346}},\ \bibinfo {pages}
  {602} (\bibinfo {year} {2014})}\BibitemShut {NoStop}%
\bibitem [{\citenamefont {Sun}\ \emph {et~al.}(2016)\citenamefont {Sun},
  \citenamefont {Zhang}, \citenamefont {Hu}, \citenamefont {Li}, \citenamefont
  {Wang}, \citenamefont {Ma}, \citenamefont {Xu}, \citenamefont {Gao},
  \citenamefont {Guan}, \citenamefont {Li}, \citenamefont {Liu}, \citenamefont
  {Qian}, \citenamefont {Zhou}, \citenamefont {Fu}, \citenamefont {Li},
  \citenamefont {Zhang},\ and\ \citenamefont {Jia}}]{Sun2016}%
  \BibitemOpen
  \bibfield  {author} {\bibinfo {author} {\bibfnamefont {H.-H.}\ \bibnamefont
  {Sun}}, \bibinfo {author} {\bibfnamefont {K.-W.}\ \bibnamefont {Zhang}},
  \bibinfo {author} {\bibfnamefont {L.-H.}\ \bibnamefont {Hu}}, \bibinfo
  {author} {\bibfnamefont {C.}~\bibnamefont {Li}}, \bibinfo {author}
  {\bibfnamefont {G.-Y.}\ \bibnamefont {Wang}}, \bibinfo {author}
  {\bibfnamefont {H.-Y.}\ \bibnamefont {Ma}}, \bibinfo {author} {\bibfnamefont
  {Z.-A.}\ \bibnamefont {Xu}}, \bibinfo {author} {\bibfnamefont {C.-L.}\
  \bibnamefont {Gao}}, \bibinfo {author} {\bibfnamefont {D.-D.}\ \bibnamefont
  {Guan}}, \bibinfo {author} {\bibfnamefont {Y.-Y.}\ \bibnamefont {Li}},
  \bibinfo {author} {\bibfnamefont {C.}~\bibnamefont {Liu}}, \bibinfo {author}
  {\bibfnamefont {D.}~\bibnamefont {Qian}}, \bibinfo {author} {\bibfnamefont
  {Y.}~\bibnamefont {Zhou}}, \bibinfo {author} {\bibfnamefont {L.}~\bibnamefont
  {Fu}}, \bibinfo {author} {\bibfnamefont {S.-C.}\ \bibnamefont {Li}}, \bibinfo
  {author} {\bibfnamefont {F.-C.}\ \bibnamefont {Zhang}}, \ and\ \bibinfo
  {author} {\bibfnamefont {J.-F.}\ \bibnamefont {Jia}},\ }\href@noop {}
  {\bibfield  {journal} {\bibinfo  {journal} {Phys. Rev. Lett.}\ }\textbf
  {\bibinfo {volume} {116}},\ \bibinfo {pages} {257003} (\bibinfo {year}
  {2016})}\BibitemShut {NoStop}%
\bibitem [{\citenamefont {Deng}\ \emph {et~al.}(2016)\citenamefont {Deng},
  \citenamefont {Vaitiek\.{e}nas}, \citenamefont {Hansen}, \citenamefont
  {Danon}, \citenamefont {Leijnse}, \citenamefont {Flensberg}, \citenamefont
  {Nyg{\aa}rd}, \citenamefont {Krogstrup},\ and\ \citenamefont
  {Marcus}}]{Deng2016}%
  \BibitemOpen
  \bibfield  {author} {\bibinfo {author} {\bibfnamefont {M.~T.}\ \bibnamefont
  {Deng}}, \bibinfo {author} {\bibfnamefont {S.}~\bibnamefont
  {Vaitiek\.{e}nas}}, \bibinfo {author} {\bibfnamefont {E.~B.}\ \bibnamefont
  {Hansen}}, \bibinfo {author} {\bibfnamefont {J.}~\bibnamefont {Danon}},
  \bibinfo {author} {\bibfnamefont {M.}~\bibnamefont {Leijnse}}, \bibinfo
  {author} {\bibfnamefont {K.}~\bibnamefont {Flensberg}}, \bibinfo {author}
  {\bibfnamefont {J.}~\bibnamefont {Nyg{\aa}rd}}, \bibinfo {author}
  {\bibfnamefont {P.}~\bibnamefont {Krogstrup}}, \ and\ \bibinfo {author}
  {\bibfnamefont {C.~M.}\ \bibnamefont {Marcus}},\ }\href@noop {} {\bibfield
  {journal} {\bibinfo  {journal} {Science}\ }\textbf {\bibinfo {volume}
  {354}},\ \bibinfo {pages} {1557} (\bibinfo {year} {2016})}\BibitemShut
  {NoStop}%
\bibitem [{\citenamefont {Zhang}\ \emph
  {et~al.}(2018{\natexlab{a}})\citenamefont {Zhang}, \citenamefont {Liu},
  \citenamefont {Gazibegovic}, \citenamefont {Xu}, \citenamefont {Logan},
  \citenamefont {Wang}, \citenamefont {van Loo}, \citenamefont {Bommer},
  \citenamefont {de~Moor}, \citenamefont {Car}, \citenamefont {Op~het Veld},
  \citenamefont {van Veldhoven}, \citenamefont {Koelling}, \citenamefont
  {Verheijen}, \citenamefont {Pendharkar}, \citenamefont {Pennachio},
  \citenamefont {Shojaei}, \citenamefont {Lee}, \citenamefont {Palmstr?m},
  \citenamefont {Bakkers}, \citenamefont {Sarma},\ and\ \citenamefont
  {Kouwenhoven}}]{ZhangH2018}%
  \BibitemOpen
  \bibfield  {author} {\bibinfo {author} {\bibfnamefont {H.}~\bibnamefont
  {Zhang}}, \bibinfo {author} {\bibfnamefont {C.-X.}\ \bibnamefont {Liu}},
  \bibinfo {author} {\bibfnamefont {S.}~\bibnamefont {Gazibegovic}}, \bibinfo
  {author} {\bibfnamefont {D.}~\bibnamefont {Xu}}, \bibinfo {author}
  {\bibfnamefont {J.~A.}\ \bibnamefont {Logan}}, \bibinfo {author}
  {\bibfnamefont {G.}~\bibnamefont {Wang}}, \bibinfo {author} {\bibfnamefont
  {N.}~\bibnamefont {van Loo}}, \bibinfo {author} {\bibfnamefont {J.~D.~S.}\
  \bibnamefont {Bommer}}, \bibinfo {author} {\bibfnamefont {M.~W.~A.}\
  \bibnamefont {de~Moor}}, \bibinfo {author} {\bibfnamefont {D.}~\bibnamefont
  {Car}}, \bibinfo {author} {\bibfnamefont {R.~L.~M.}\ \bibnamefont {Op~het
  Veld}}, \bibinfo {author} {\bibfnamefont {P.~J.}\ \bibnamefont {van
  Veldhoven}}, \bibinfo {author} {\bibfnamefont {S.}~\bibnamefont {Koelling}},
  \bibinfo {author} {\bibfnamefont {M.~A.}\ \bibnamefont {Verheijen}}, \bibinfo
  {author} {\bibfnamefont {M.}~\bibnamefont {Pendharkar}}, \bibinfo {author}
  {\bibfnamefont {D.~J.}\ \bibnamefont {Pennachio}}, \bibinfo {author}
  {\bibfnamefont {B.}~\bibnamefont {Shojaei}}, \bibinfo {author} {\bibfnamefont
  {J.~S.}\ \bibnamefont {Lee}}, \bibinfo {author} {\bibfnamefont {C.~J.}\
  \bibnamefont {Palmstr?m}}, \bibinfo {author} {\bibfnamefont {E.~P. A.~M.}\
  \bibnamefont {Bakkers}}, \bibinfo {author} {\bibfnamefont {S.~D.}\
  \bibnamefont {Sarma}}, \ and\ \bibinfo {author} {\bibfnamefont {L.~P.}\
  \bibnamefont {Kouwenhoven}},\ }\href@noop {} {\bibfield  {journal} {\bibinfo
  {journal} {Nature}\ }\textbf {\bibinfo {volume} {556}},\ \bibinfo {pages}
  {74} (\bibinfo {year} {2018}{\natexlab{a}})}\BibitemShut {NoStop}%
\bibitem [{\citenamefont {Lutchyn}\ \emph {et~al.}(2018)\citenamefont
  {Lutchyn}, \citenamefont {Bakkers}, \citenamefont {Kouwenhoven},
  \citenamefont {Krogstrup}, \citenamefont {Marcus},\ and\ \citenamefont
  {Oreg}}]{Lutchyn2018}%
  \BibitemOpen
  \bibfield  {author} {\bibinfo {author} {\bibfnamefont {R.~M.}\ \bibnamefont
  {Lutchyn}}, \bibinfo {author} {\bibfnamefont {E.~P. A.~M.}\ \bibnamefont
  {Bakkers}}, \bibinfo {author} {\bibfnamefont {L.~P.}\ \bibnamefont
  {Kouwenhoven}}, \bibinfo {author} {\bibfnamefont {P.}~\bibnamefont
  {Krogstrup}}, \bibinfo {author} {\bibfnamefont {C.~M.}\ \bibnamefont
  {Marcus}}, \ and\ \bibinfo {author} {\bibfnamefont {Y.}~\bibnamefont
  {Oreg}},\ }\href@noop {} {\bibfield  {journal} {\bibinfo  {journal} {Nat.
  Rev. Mater.}\ }\textbf {\bibinfo {volume} {3}},\ \bibinfo {pages} {52}
  (\bibinfo {year} {2018})}\BibitemShut {NoStop}%
\bibitem [{\citenamefont {Hao}\ and\ \citenamefont {Hu}(2014)}]{Hao2014}%
  \BibitemOpen
  \bibfield  {author} {\bibinfo {author} {\bibfnamefont {N.}~\bibnamefont
  {Hao}}\ and\ \bibinfo {author} {\bibfnamefont {J.}~\bibnamefont {Hu}},\
  }\href@noop {} {\bibfield  {journal} {\bibinfo  {journal} {Phys. Rev. X}\
  }\textbf {\bibinfo {volume} {4}},\ \bibinfo {pages} {031053} (\bibinfo {year}
  {2014})}\BibitemShut {NoStop}%
\bibitem [{\citenamefont {Wu}\ \emph {et~al.}(2016)\citenamefont {Wu},
  \citenamefont {Qin}, \citenamefont {Liang}, \citenamefont {Fan},\ and\
  \citenamefont {Hu}}]{Wu2016}%
  \BibitemOpen
  \bibfield  {author} {\bibinfo {author} {\bibfnamefont {X.}~\bibnamefont
  {Wu}}, \bibinfo {author} {\bibfnamefont {S.}~\bibnamefont {Qin}}, \bibinfo
  {author} {\bibfnamefont {Y.}~\bibnamefont {Liang}}, \bibinfo {author}
  {\bibfnamefont {H.}~\bibnamefont {Fan}}, \ and\ \bibinfo {author}
  {\bibfnamefont {J.}~\bibnamefont {Hu}},\ }\href@noop {} {\bibfield  {journal}
  {\bibinfo  {journal} {Phys. Rev. B}\ }\textbf {\bibinfo {volume} {93}},\
  \bibinfo {pages} {115129} (\bibinfo {year} {2016})}\BibitemShut {NoStop}%
\bibitem [{\citenamefont {Wang}\ \emph {et~al.}(2015)\citenamefont {Wang},
  \citenamefont {Zhang}, \citenamefont {Xu}, \citenamefont {Zeng},
  \citenamefont {Miao}, \citenamefont {Xu}, \citenamefont {Qian}, \citenamefont
  {Weng}, \citenamefont {Richard}, \citenamefont {Fedorov}, \citenamefont
  {Ding}, \citenamefont {Dai},\ and\ \citenamefont {Fang}}]{Wang2015}%
  \BibitemOpen
  \bibfield  {author} {\bibinfo {author} {\bibfnamefont {Z.}~\bibnamefont
  {Wang}}, \bibinfo {author} {\bibfnamefont {P.}~\bibnamefont {Zhang}},
  \bibinfo {author} {\bibfnamefont {G.}~\bibnamefont {Xu}}, \bibinfo {author}
  {\bibfnamefont {L.~K.}\ \bibnamefont {Zeng}}, \bibinfo {author}
  {\bibfnamefont {H.}~\bibnamefont {Miao}}, \bibinfo {author} {\bibfnamefont
  {X.}~\bibnamefont {Xu}}, \bibinfo {author} {\bibfnamefont {T.}~\bibnamefont
  {Qian}}, \bibinfo {author} {\bibfnamefont {H.}~\bibnamefont {Weng}}, \bibinfo
  {author} {\bibfnamefont {P.}~\bibnamefont {Richard}}, \bibinfo {author}
  {\bibfnamefont {A.~V.}\ \bibnamefont {Fedorov}}, \bibinfo {author}
  {\bibfnamefont {H.}~\bibnamefont {Ding}}, \bibinfo {author} {\bibfnamefont
  {X.}~\bibnamefont {Dai}}, \ and\ \bibinfo {author} {\bibfnamefont
  {Z.}~\bibnamefont {Fang}},\ }\href@noop {} {\bibfield  {journal} {\bibinfo
  {journal} {Phys. Rev. B}\ }\textbf {\bibinfo {volume} {92}},\ \bibinfo
  {pages} {115119} (\bibinfo {year} {2015})}\BibitemShut {NoStop}%
\bibitem [{\citenamefont {Xu}\ \emph {et~al.}(2016)\citenamefont {Xu},
  \citenamefont {Lian}, \citenamefont {Tang}, \citenamefont {Qi},\ and\
  \citenamefont {Zhang}}]{XuPRL2016}%
  \BibitemOpen
  \bibfield  {author} {\bibinfo {author} {\bibfnamefont {G.}~\bibnamefont
  {Xu}}, \bibinfo {author} {\bibfnamefont {B.}~\bibnamefont {Lian}}, \bibinfo
  {author} {\bibfnamefont {P.}~\bibnamefont {Tang}}, \bibinfo {author}
  {\bibfnamefont {X.-L.}\ \bibnamefont {Qi}}, \ and\ \bibinfo {author}
  {\bibfnamefont {S.-C.}\ \bibnamefont {Zhang}},\ }\href@noop {} {\bibfield
  {journal} {\bibinfo  {journal} {Phys. Rev. Lett.}\ }\textbf {\bibinfo
  {volume} {117}},\ \bibinfo {pages} {047001} (\bibinfo {year}
  {2016})}\BibitemShut {NoStop}%
\bibitem [{\citenamefont {Zhang}\ \emph
  {et~al.}(2018{\natexlab{b}})\citenamefont {Zhang}, \citenamefont {Yaji},
  \citenamefont {Hashimoto}, \citenamefont {Ota}, \citenamefont {Kondo},
  \citenamefont {Okazaki}, \citenamefont {Wang}, \citenamefont {Wen},
  \citenamefont {Gu}, \citenamefont {Ding},\ and\ \citenamefont
  {Shin}}]{Zhang2018}%
  \BibitemOpen
  \bibfield  {author} {\bibinfo {author} {\bibfnamefont {P.}~\bibnamefont
  {Zhang}}, \bibinfo {author} {\bibfnamefont {K.}~\bibnamefont {Yaji}},
  \bibinfo {author} {\bibfnamefont {T.}~\bibnamefont {Hashimoto}}, \bibinfo
  {author} {\bibfnamefont {Y.}~\bibnamefont {Ota}}, \bibinfo {author}
  {\bibfnamefont {T.}~\bibnamefont {Kondo}}, \bibinfo {author} {\bibfnamefont
  {K.}~\bibnamefont {Okazaki}}, \bibinfo {author} {\bibfnamefont
  {Z.}~\bibnamefont {Wang}}, \bibinfo {author} {\bibfnamefont {J.}~\bibnamefont
  {Wen}}, \bibinfo {author} {\bibfnamefont {G.~D.}\ \bibnamefont {Gu}},
  \bibinfo {author} {\bibfnamefont {H.}~\bibnamefont {Ding}}, \ and\ \bibinfo
  {author} {\bibfnamefont {S.}~\bibnamefont {Shin}},\ }\href@noop {} {\bibfield
   {journal} {\bibinfo  {journal} {Science}\ }\textbf {\bibinfo {volume}
  {360}},\ \bibinfo {pages} {182} (\bibinfo {year}
  {2018}{\natexlab{b}})}\BibitemShut {NoStop}%
\bibitem [{\citenamefont {Zhang}\ \emph
  {et~al.}(2019{\natexlab{a}})\citenamefont {Zhang}, \citenamefont {Wang},
  \citenamefont {Wu}, \citenamefont {Yaji}, \citenamefont {Ishida},
  \citenamefont {Kohama}, \citenamefont {Dai}, \citenamefont {Sun},
  \citenamefont {Bareille}, \citenamefont {Kuroda}, \citenamefont {Kondo},
  \citenamefont {Okazaki}, \citenamefont {Kindo}, \citenamefont {Wang},
  \citenamefont {Jin}, \citenamefont {Hu}, \citenamefont {Thomale},
  \citenamefont {Sumida}, \citenamefont {Wu}, \citenamefont {Miyamoto},
  \citenamefont {Okuda}, \citenamefont {Ding}, \citenamefont {Gu},
  \citenamefont {Tamegai}, \citenamefont {Kawakami}, \citenamefont {Sato},\
  and\ \citenamefont {Shin}}]{Zhang2019NP}%
  \BibitemOpen
  \bibfield  {author} {\bibinfo {author} {\bibfnamefont {P.}~\bibnamefont
  {Zhang}}, \bibinfo {author} {\bibfnamefont {Z.}~\bibnamefont {Wang}},
  \bibinfo {author} {\bibfnamefont {X.}~\bibnamefont {Wu}}, \bibinfo {author}
  {\bibfnamefont {K.}~\bibnamefont {Yaji}}, \bibinfo {author} {\bibfnamefont
  {Y.}~\bibnamefont {Ishida}}, \bibinfo {author} {\bibfnamefont
  {Y.}~\bibnamefont {Kohama}}, \bibinfo {author} {\bibfnamefont
  {G.}~\bibnamefont {Dai}}, \bibinfo {author} {\bibfnamefont {Y.}~\bibnamefont
  {Sun}}, \bibinfo {author} {\bibfnamefont {C.}~\bibnamefont {Bareille}},
  \bibinfo {author} {\bibfnamefont {K.}~\bibnamefont {Kuroda}}, \bibinfo
  {author} {\bibfnamefont {T.}~\bibnamefont {Kondo}}, \bibinfo {author}
  {\bibfnamefont {K.}~\bibnamefont {Okazaki}}, \bibinfo {author} {\bibfnamefont
  {K.}~\bibnamefont {Kindo}}, \bibinfo {author} {\bibfnamefont
  {X.}~\bibnamefont {Wang}}, \bibinfo {author} {\bibfnamefont {C.}~\bibnamefont
  {Jin}}, \bibinfo {author} {\bibfnamefont {J.}~\bibnamefont {Hu}}, \bibinfo
  {author} {\bibfnamefont {R.}~\bibnamefont {Thomale}}, \bibinfo {author}
  {\bibfnamefont {K.}~\bibnamefont {Sumida}}, \bibinfo {author} {\bibfnamefont
  {S.}~\bibnamefont {Wu}}, \bibinfo {author} {\bibfnamefont {K.}~\bibnamefont
  {Miyamoto}}, \bibinfo {author} {\bibfnamefont {T.}~\bibnamefont {Okuda}},
  \bibinfo {author} {\bibfnamefont {H.}~\bibnamefont {Ding}}, \bibinfo {author}
  {\bibfnamefont {G.~D.}\ \bibnamefont {Gu}}, \bibinfo {author} {\bibfnamefont
  {T.}~\bibnamefont {Tamegai}}, \bibinfo {author} {\bibfnamefont
  {T.}~\bibnamefont {Kawakami}}, \bibinfo {author} {\bibfnamefont
  {M.}~\bibnamefont {Sato}}, \ and\ \bibinfo {author} {\bibfnamefont
  {S.}~\bibnamefont {Shin}},\ }\href@noop {} {\bibfield  {journal} {\bibinfo
  {journal} {Nat. Phys.}\ }\textbf {\bibinfo {volume} {15}},\ \bibinfo {pages}
  {41} (\bibinfo {year} {2019}{\natexlab{a}})}\BibitemShut {NoStop}%
\bibitem [{\citenamefont {Wang}\ \emph
  {et~al.}(2018{\natexlab{a}})\citenamefont {Wang}, \citenamefont {Kong},
  \citenamefont {Fan}, \citenamefont {Chen}, \citenamefont {Zhu}, \citenamefont
  {Liu}, \citenamefont {Cao}, \citenamefont {Sun}, \citenamefont {Du},
  \citenamefont {Schneeloch}, \citenamefont {Zhong}, \citenamefont {Gu},
  \citenamefont {Fu}, \citenamefont {Ding},\ and\ \citenamefont
  {Gao}}]{WangDF2018}%
  \BibitemOpen
  \bibfield  {author} {\bibinfo {author} {\bibfnamefont {D.}~\bibnamefont
  {Wang}}, \bibinfo {author} {\bibfnamefont {L.}~\bibnamefont {Kong}}, \bibinfo
  {author} {\bibfnamefont {P.}~\bibnamefont {Fan}}, \bibinfo {author}
  {\bibfnamefont {H.}~\bibnamefont {Chen}}, \bibinfo {author} {\bibfnamefont
  {S.}~\bibnamefont {Zhu}}, \bibinfo {author} {\bibfnamefont {W.}~\bibnamefont
  {Liu}}, \bibinfo {author} {\bibfnamefont {L.}~\bibnamefont {Cao}}, \bibinfo
  {author} {\bibfnamefont {Y.}~\bibnamefont {Sun}}, \bibinfo {author}
  {\bibfnamefont {S.}~\bibnamefont {Du}}, \bibinfo {author} {\bibfnamefont
  {J.}~\bibnamefont {Schneeloch}}, \bibinfo {author} {\bibfnamefont
  {R.}~\bibnamefont {Zhong}}, \bibinfo {author} {\bibfnamefont
  {G.}~\bibnamefont {Gu}}, \bibinfo {author} {\bibfnamefont {L.}~\bibnamefont
  {Fu}}, \bibinfo {author} {\bibfnamefont {H.}~\bibnamefont {Ding}}, \ and\
  \bibinfo {author} {\bibfnamefont {H.-J.}\ \bibnamefont {Gao}},\ }\href@noop
  {} {\bibfield  {journal} {\bibinfo  {journal} {Science}\ }\textbf {\bibinfo
  {volume} {362}},\ \bibinfo {pages} {333} (\bibinfo {year}
  {2018}{\natexlab{a}})}\BibitemShut {NoStop}%
\bibitem [{\citenamefont {Shi}\ \emph {et~al.}(2017)\citenamefont {Shi},
  \citenamefont {Han}, \citenamefont {Richard}, \citenamefont {Wu},
  \citenamefont {Peng}, \citenamefont {Qian}, \citenamefont {Wang},
  \citenamefont {Hu}, \citenamefont {Sun},\ and\ \citenamefont
  {Ding}}]{Shi2017}%
  \BibitemOpen
  \bibfield  {author} {\bibinfo {author} {\bibfnamefont {X.}~\bibnamefont
  {Shi}}, \bibinfo {author} {\bibfnamefont {Z.-Q.}\ \bibnamefont {Han}},
  \bibinfo {author} {\bibfnamefont {P.}~\bibnamefont {Richard}}, \bibinfo
  {author} {\bibfnamefont {X.-X.}\ \bibnamefont {Wu}}, \bibinfo {author}
  {\bibfnamefont {X.-L.}\ \bibnamefont {Peng}}, \bibinfo {author}
  {\bibfnamefont {T.}~\bibnamefont {Qian}}, \bibinfo {author} {\bibfnamefont
  {S.-C.}\ \bibnamefont {Wang}}, \bibinfo {author} {\bibfnamefont {J.-P.}\
  \bibnamefont {Hu}}, \bibinfo {author} {\bibfnamefont {Y.-J.}\ \bibnamefont
  {Sun}}, \ and\ \bibinfo {author} {\bibfnamefont {H.}~\bibnamefont {Ding}},\
  }\href@noop {} {\bibfield  {journal} {\bibinfo  {journal} {Sci. Bull.}\
  }\textbf {\bibinfo {volume} {62}},\ \bibinfo {pages} {503} (\bibinfo {year}
  {2017})}\BibitemShut {NoStop}%
\bibitem [{\citenamefont {Li}\ \emph {et~al.}(2015)\citenamefont {Li},
  \citenamefont {Ding}, \citenamefont {Tang}, \citenamefont {Peng},
  \citenamefont {Zhang}, \citenamefont {Zhang}, \citenamefont {Zhou},
  \citenamefont {Zhang}, \citenamefont {Song}, \citenamefont {He},
  \citenamefont {Ji}, \citenamefont {Chen}, \citenamefont {Gu}, \citenamefont
  {Wang}, \citenamefont {Ma},\ and\ \citenamefont {Xue}}]{Li2015PRB}%
  \BibitemOpen
  \bibfield  {author} {\bibinfo {author} {\bibfnamefont {F.}~\bibnamefont
  {Li}}, \bibinfo {author} {\bibfnamefont {H.}~\bibnamefont {Ding}}, \bibinfo
  {author} {\bibfnamefont {C.}~\bibnamefont {Tang}}, \bibinfo {author}
  {\bibfnamefont {J.}~\bibnamefont {Peng}}, \bibinfo {author} {\bibfnamefont
  {Q.}~\bibnamefont {Zhang}}, \bibinfo {author} {\bibfnamefont
  {W.}~\bibnamefont {Zhang}}, \bibinfo {author} {\bibfnamefont
  {G.}~\bibnamefont {Zhou}}, \bibinfo {author} {\bibfnamefont {D.}~\bibnamefont
  {Zhang}}, \bibinfo {author} {\bibfnamefont {C.-L.}\ \bibnamefont {Song}},
  \bibinfo {author} {\bibfnamefont {K.}~\bibnamefont {He}}, \bibinfo {author}
  {\bibfnamefont {S.}~\bibnamefont {Ji}}, \bibinfo {author} {\bibfnamefont
  {X.}~\bibnamefont {Chen}}, \bibinfo {author} {\bibfnamefont {L.}~\bibnamefont
  {Gu}}, \bibinfo {author} {\bibfnamefont {L.}~\bibnamefont {Wang}}, \bibinfo
  {author} {\bibfnamefont {X.-C.}\ \bibnamefont {Ma}}, \ and\ \bibinfo {author}
  {\bibfnamefont {Q.-K.}\ \bibnamefont {Xue}},\ }\href@noop {} {\bibfield
  {journal} {\bibinfo  {journal} {Phys. Rev. B}\ }\textbf {\bibinfo {volume}
  {91}},\ \bibinfo {pages} {220503} (\bibinfo {year} {2015})}\BibitemShut
  {NoStop}%
\bibitem [{\citenamefont {Salamon}\ \emph {et~al.}(2016)\citenamefont
  {Salamon}, \citenamefont {Cornell}, \citenamefont {Jaime}, \citenamefont
  {Balakirev}, \citenamefont {Zakhidov}, \citenamefont {Huang},\ and\
  \citenamefont {Wang}}]{Salamon2016}%
  \BibitemOpen
  \bibfield  {author} {\bibinfo {author} {\bibfnamefont {M.~B.}\ \bibnamefont
  {Salamon}}, \bibinfo {author} {\bibfnamefont {N.}~\bibnamefont {Cornell}},
  \bibinfo {author} {\bibfnamefont {M.}~\bibnamefont {Jaime}}, \bibinfo
  {author} {\bibfnamefont {F.~F.}\ \bibnamefont {Balakirev}}, \bibinfo {author}
  {\bibfnamefont {A.}~\bibnamefont {Zakhidov}}, \bibinfo {author}
  {\bibfnamefont {J.}~\bibnamefont {Huang}}, \ and\ \bibinfo {author}
  {\bibfnamefont {H.}~\bibnamefont {Wang}},\ }\href@noop {} {\bibfield
  {journal} {\bibinfo  {journal} {Sci. Rep.}\ }\textbf {\bibinfo {volume}
  {6}},\ \bibinfo {pages} {21469} (\bibinfo {year} {2016})}\BibitemShut
  {NoStop}%
\bibitem [{\citenamefont {Hirschfeld}\ \emph {et~al.}(2011)\citenamefont
  {Hirschfeld}, \citenamefont {Korshunov},\ and\ \citenamefont
  {Mazin}}]{Hirschfeld2011}%
  \BibitemOpen
  \bibfield  {author} {\bibinfo {author} {\bibfnamefont {P.~J.}\ \bibnamefont
  {Hirschfeld}}, \bibinfo {author} {\bibfnamefont {M.~M.}\ \bibnamefont
  {Korshunov}}, \ and\ \bibinfo {author} {\bibfnamefont {I.~I.}\ \bibnamefont
  {Mazin}},\ }\href@noop {} {\bibfield  {journal} {\bibinfo  {journal} {Rep.
  Prog. Phys.}\ }\textbf {\bibinfo {volume} {74}},\ \bibinfo {pages} {124508}
  (\bibinfo {year} {2011})}\BibitemShut {NoStop}%
\bibitem [{\citenamefont {Nourafkan}\ \emph {et~al.}(2016)\citenamefont
  {Nourafkan}, \citenamefont {Kotliar},\ and\ \citenamefont
  {Tremblay}}]{Nourafkan2016}%
  \BibitemOpen
  \bibfield  {author} {\bibinfo {author} {\bibfnamefont {R.}~\bibnamefont
  {Nourafkan}}, \bibinfo {author} {\bibfnamefont {G.}~\bibnamefont {Kotliar}},
  \ and\ \bibinfo {author} {\bibfnamefont {A.~M.~S.}\ \bibnamefont
  {Tremblay}},\ }\href@noop {} {\bibfield  {journal} {\bibinfo  {journal}
  {Phys. Rev. Lett.}\ }\textbf {\bibinfo {volume} {117}},\ \bibinfo {pages}
  {137001} (\bibinfo {year} {2016})}\BibitemShut {NoStop}%
\bibitem [{\citenamefont {Douglass}(1961)}]{Douglass1961}%
  \BibitemOpen
  \bibfield  {author} {\bibinfo {author} {\bibfnamefont {D.~H.}\ \bibnamefont
  {Douglass}},\ }\href@noop {} {\bibfield  {journal} {\bibinfo  {journal}
  {Phys. Rev. Lett.}\ }\textbf {\bibinfo {volume} {6}},\ \bibinfo {pages} {346}
  (\bibinfo {year} {1961})}\BibitemShut {NoStop}%
\bibitem [{\citenamefont {Bernevig}\ \emph {et~al.}(2006)\citenamefont
  {Bernevig}, \citenamefont {Hughes},\ and\ \citenamefont
  {Zhang}}]{Bernevig2006}%
  \BibitemOpen
  \bibfield  {author} {\bibinfo {author} {\bibfnamefont {B.~A.}\ \bibnamefont
  {Bernevig}}, \bibinfo {author} {\bibfnamefont {T.~L.}\ \bibnamefont
  {Hughes}}, \ and\ \bibinfo {author} {\bibfnamefont {S.-C.}\ \bibnamefont
  {Zhang}},\ }\href@noop {} {\bibfield  {journal} {\bibinfo  {journal}
  {Science}\ }\textbf {\bibinfo {volume} {314}},\ \bibinfo {pages} {1757}
  (\bibinfo {year} {2006})}\BibitemShut {NoStop}%
\bibitem [{\citenamefont {Yan}\ \emph {et~al.}(2018)\citenamefont {Yan},
  \citenamefont {Song},\ and\ \citenamefont {Wang}}]{YanZB2018}%
  \BibitemOpen
  \bibfield  {author} {\bibinfo {author} {\bibfnamefont {Z.}~\bibnamefont
  {Yan}}, \bibinfo {author} {\bibfnamefont {F.}~\bibnamefont {Song}}, \ and\
  \bibinfo {author} {\bibfnamefont {Z.}~\bibnamefont {Wang}},\ }\href@noop {}
  {\bibfield  {journal} {\bibinfo  {journal} {Phys. Rev. Lett.}\ }\textbf
  {\bibinfo {volume} {121}},\ \bibinfo {pages} {096803} (\bibinfo {year}
  {2018})}\BibitemShut {NoStop}%
\bibitem [{\citenamefont {Wang}\ \emph
  {et~al.}(2018{\natexlab{b}})\citenamefont {Wang}, \citenamefont {Liu},
  \citenamefont {Lu},\ and\ \citenamefont {Zhang}}]{WangQY2018}%
  \BibitemOpen
  \bibfield  {author} {\bibinfo {author} {\bibfnamefont {Q.}~\bibnamefont
  {Wang}}, \bibinfo {author} {\bibfnamefont {C.-C.}\ \bibnamefont {Liu}},
  \bibinfo {author} {\bibfnamefont {Y.-M.}\ \bibnamefont {Lu}}, \ and\ \bibinfo
  {author} {\bibfnamefont {F.}~\bibnamefont {Zhang}},\ }\href@noop {}
  {\bibfield  {journal} {\bibinfo  {journal} {Phys. Rev. Lett.}\ }\textbf
  {\bibinfo {volume} {121}},\ \bibinfo {pages} {186801} (\bibinfo {year}
  {2018}{\natexlab{b}})}\BibitemShut {NoStop}%
\bibitem [{\citenamefont {Wang}\ \emph
  {et~al.}(2018{\natexlab{c}})\citenamefont {Wang}, \citenamefont {Lin},\ and\
  \citenamefont {Hughes}}]{WangYX2018}%
  \BibitemOpen
  \bibfield  {author} {\bibinfo {author} {\bibfnamefont {Y.}~\bibnamefont
  {Wang}}, \bibinfo {author} {\bibfnamefont {M.}~\bibnamefont {Lin}}, \ and\
  \bibinfo {author} {\bibfnamefont {T.~L.}\ \bibnamefont {Hughes}},\
  }\href@noop {} {\bibfield  {journal} {\bibinfo  {journal} {Phys. Rev. B}\
  }\textbf {\bibinfo {volume} {98}},\ \bibinfo {pages} {165144} (\bibinfo
  {year} {2018}{\natexlab{c}})}\BibitemShut {NoStop}%
\bibitem [{\citenamefont {Pan}\ \emph {et~al.}(2019)\citenamefont {Pan},
  \citenamefont {Yang}, \citenamefont {Chen}, \citenamefont {Xu}, \citenamefont
  {Liu},\ and\ \citenamefont {Liu}}]{pan2018lattice}%
  \BibitemOpen
  \bibfield  {author} {\bibinfo {author} {\bibfnamefont {X.-H.}\ \bibnamefont
  {Pan}}, \bibinfo {author} {\bibfnamefont {K.-J.}\ \bibnamefont {Yang}},
  \bibinfo {author} {\bibfnamefont {L.}~\bibnamefont {Chen}}, \bibinfo {author}
  {\bibfnamefont {G.}~\bibnamefont {Xu}}, \bibinfo {author} {\bibfnamefont
  {C.-X.}\ \bibnamefont {Liu}}, \ and\ \bibinfo {author} {\bibfnamefont
  {X.}~\bibnamefont {Liu}},\ }\href@noop {} {\bibfield  {journal} {\bibinfo
  {journal} {Phys. Rev. Lett.}\ }\textbf {\bibinfo {volume} {123}},\ \bibinfo
  {pages} {156801} (\bibinfo {year} {2019})}\BibitemShut {NoStop}%
\bibitem [{\citenamefont {Zhu}(2018)}]{zhu2018tunable}%
  \BibitemOpen
  \bibfield  {author} {\bibinfo {author} {\bibfnamefont {X.}~\bibnamefont
  {Zhu}},\ }\href@noop {} {\bibfield  {journal} {\bibinfo  {journal} {Phys.
  Rev. B}\ }\textbf {\bibinfo {volume} {97}},\ \bibinfo {pages} {205134}
  (\bibinfo {year} {2018})}\BibitemShut {NoStop}%
\bibitem [{\citenamefont {Geier}\ \emph {et~al.}(2018)\citenamefont {Geier},
  \citenamefont {Trifunovic}, \citenamefont {Hoskam},\ and\ \citenamefont
  {Brouwer}}]{geier2018second}%
  \BibitemOpen
  \bibfield  {author} {\bibinfo {author} {\bibfnamefont {M.}~\bibnamefont
  {Geier}}, \bibinfo {author} {\bibfnamefont {L.}~\bibnamefont {Trifunovic}},
  \bibinfo {author} {\bibfnamefont {M.}~\bibnamefont {Hoskam}}, \ and\ \bibinfo
  {author} {\bibfnamefont {P.~W.}\ \bibnamefont {Brouwer}},\ }\href@noop {}
  {\bibfield  {journal} {\bibinfo  {journal} {Phys. Rev. B}\ }\textbf {\bibinfo
  {volume} {97}},\ \bibinfo {pages} {205135} (\bibinfo {year}
  {2018})}\BibitemShut {NoStop}%
\bibitem [{\citenamefont {Khalaf}(2018)}]{khalaf2018higher}%
  \BibitemOpen
  \bibfield  {author} {\bibinfo {author} {\bibfnamefont {E.}~\bibnamefont
  {Khalaf}},\ }\href@noop {} {\bibfield  {journal} {\bibinfo  {journal} {Phys.
  Rev. B}\ }\textbf {\bibinfo {volume} {97}},\ \bibinfo {pages} {205136}
  (\bibinfo {year} {2018})}\BibitemShut {NoStop}%
\bibitem [{\citenamefont {Wu}\ \emph {et~al.}(2019{\natexlab{a}})\citenamefont
  {Wu}, \citenamefont {Yan},\ and\ \citenamefont {Huang}}]{wu2019higher}%
  \BibitemOpen
  \bibfield  {author} {\bibinfo {author} {\bibfnamefont {Z.}~\bibnamefont
  {Wu}}, \bibinfo {author} {\bibfnamefont {Z.}~\bibnamefont {Yan}}, \ and\
  \bibinfo {author} {\bibfnamefont {W.}~\bibnamefont {Huang}},\ }\href@noop {}
  {\bibfield  {journal} {\bibinfo  {journal} {Phys. Rev. B}\ }\textbf {\bibinfo
  {volume} {99}},\ \bibinfo {pages} {020508} (\bibinfo {year}
  {2019}{\natexlab{a}})}\BibitemShut {NoStop}%
\bibitem [{\citenamefont {Zhang}\ \emph
  {et~al.}(2019{\natexlab{b}})\citenamefont {Zhang}, \citenamefont {Cole},\
  and\ \citenamefont {Das~Sarma}}]{ZhangRX2019PRL}%
  \BibitemOpen
  \bibfield  {author} {\bibinfo {author} {\bibfnamefont {R.-X.}\ \bibnamefont
  {Zhang}}, \bibinfo {author} {\bibfnamefont {W.~S.}\ \bibnamefont {Cole}}, \
  and\ \bibinfo {author} {\bibfnamefont {S.}~\bibnamefont {Das~Sarma}},\
  }\href@noop {} {\bibfield  {journal} {\bibinfo  {journal} {Phys. Rev. Lett.}\
  }\textbf {\bibinfo {volume} {122}},\ \bibinfo {pages} {187001} (\bibinfo
  {year} {2019}{\natexlab{b}})}\BibitemShut {NoStop}%
\bibitem [{\citenamefont {Volpez}\ \emph {et~al.}(2019)\citenamefont {Volpez},
  \citenamefont {Loss},\ and\ \citenamefont {Klinovaja}}]{Volpez2019}%
  \BibitemOpen
  \bibfield  {author} {\bibinfo {author} {\bibfnamefont {Y.}~\bibnamefont
  {Volpez}}, \bibinfo {author} {\bibfnamefont {D.}~\bibnamefont {Loss}}, \ and\
  \bibinfo {author} {\bibfnamefont {J.}~\bibnamefont {Klinovaja}},\ }\href@noop
  {} {\bibfield  {journal} {\bibinfo  {journal} {Phys. Rev. Lett.}\ }\textbf
  {\bibinfo {volume} {122}},\ \bibinfo {pages} {126402} (\bibinfo {year}
  {2019})}\BibitemShut {NoStop}%
\bibitem [{\citenamefont {Ghorashi}\ \emph {et~al.}(2019)\citenamefont
  {Ghorashi}, \citenamefont {Hu}, \citenamefont {Hughes},\ and\ \citenamefont
  {Rossi}}]{Ghorashi2019}%
  \BibitemOpen
  \bibfield  {author} {\bibinfo {author} {\bibfnamefont {S.~A.~A.}\
  \bibnamefont {Ghorashi}}, \bibinfo {author} {\bibfnamefont {X.}~\bibnamefont
  {Hu}}, \bibinfo {author} {\bibfnamefont {T.~L.}\ \bibnamefont {Hughes}}, \
  and\ \bibinfo {author} {\bibfnamefont {E.}~\bibnamefont {Rossi}},\
  }\href@noop {} {\bibfield  {journal} {\bibinfo  {journal} {arXiv:1901.07579}\
  } (\bibinfo {year} {2019})}\BibitemShut {NoStop}%
\bibitem [{\citenamefont {Mostofi}\ \emph {et~al.}(2014)\citenamefont
  {Mostofi}, \citenamefont {Yates}, \citenamefont {Pizzi}, \citenamefont {Lee},
  \citenamefont {Souza}, \citenamefont {Vanderbilt},\ and\ \citenamefont
  {Marzari}}]{Mostofi2014}%
  \BibitemOpen
  \bibfield  {author} {\bibinfo {author} {\bibfnamefont {A.~A.}\ \bibnamefont
  {Mostofi}}, \bibinfo {author} {\bibfnamefont {J.~R.}\ \bibnamefont {Yates}},
  \bibinfo {author} {\bibfnamefont {G.}~\bibnamefont {Pizzi}}, \bibinfo
  {author} {\bibfnamefont {Y.-S.}\ \bibnamefont {Lee}}, \bibinfo {author}
  {\bibfnamefont {I.}~\bibnamefont {Souza}}, \bibinfo {author} {\bibfnamefont
  {D.}~\bibnamefont {Vanderbilt}}, \ and\ \bibinfo {author} {\bibfnamefont
  {N.}~\bibnamefont {Marzari}},\ }\href@noop {} {\bibfield  {journal} {\bibinfo
   {journal} {Comput. Phys. Commun.}\ }\textbf {\bibinfo {volume} {185}},\
  \bibinfo {pages} {2309} (\bibinfo {year} {2014})}\BibitemShut {NoStop}%
\bibitem [{\citenamefont {Huang}\ \emph {et~al.}(2018)\citenamefont {Huang},
  \citenamefont {Clark}, \citenamefont {Klein}, \citenamefont {MacNeill},
  \citenamefont {Navarro-Moratalla}, \citenamefont {Seyler}, \citenamefont
  {Wilson}, \citenamefont {McGuire}, \citenamefont {Cobden}, \citenamefont
  {Xiao}, \citenamefont {Yao}, \citenamefont {Jarillo-Herrero},\ and\
  \citenamefont {Xu}}]{Huang2018}%
  \BibitemOpen
  \bibfield  {author} {\bibinfo {author} {\bibfnamefont {B.}~\bibnamefont
  {Huang}}, \bibinfo {author} {\bibfnamefont {G.}~\bibnamefont {Clark}},
  \bibinfo {author} {\bibfnamefont {D.~R.}\ \bibnamefont {Klein}}, \bibinfo
  {author} {\bibfnamefont {D.}~\bibnamefont {MacNeill}}, \bibinfo {author}
  {\bibfnamefont {E.}~\bibnamefont {Navarro-Moratalla}}, \bibinfo {author}
  {\bibfnamefont {K.~L.}\ \bibnamefont {Seyler}}, \bibinfo {author}
  {\bibfnamefont {N.}~\bibnamefont {Wilson}}, \bibinfo {author} {\bibfnamefont
  {M.~A.}\ \bibnamefont {McGuire}}, \bibinfo {author} {\bibfnamefont {D.~H.}\
  \bibnamefont {Cobden}}, \bibinfo {author} {\bibfnamefont {D.}~\bibnamefont
  {Xiao}}, \bibinfo {author} {\bibfnamefont {W.}~\bibnamefont {Yao}}, \bibinfo
  {author} {\bibfnamefont {P.}~\bibnamefont {Jarillo-Herrero}}, \ and\ \bibinfo
  {author} {\bibfnamefont {X.}~\bibnamefont {Xu}},\ }\href@noop {} {\bibfield
  {journal} {\bibinfo  {journal} {Nat. Nanotechnol.}\ }\textbf {\bibinfo
  {volume} {13}},\ \bibinfo {pages} {544} (\bibinfo {year} {2018})}\BibitemShut
  {NoStop}%
\bibitem [{\citenamefont {Ghazaryan}\ \emph {et~al.}(2018)\citenamefont
  {Ghazaryan}, \citenamefont {Greenaway}, \citenamefont {Wang}, \citenamefont
  {Guarochico-Moreira}, \citenamefont {Vera-Marun}, \citenamefont {Yin},
  \citenamefont {Liao}, \citenamefont {Morozov}, \citenamefont {Kristanovski},
  \citenamefont {Lichtenstein}, \citenamefont {Katsnelson}, \citenamefont
  {Withers}, \citenamefont {Mishchenko}, \citenamefont {Eaves}, \citenamefont
  {Geim}, \citenamefont {Novoselov},\ and\ \citenamefont
  {Misra}}]{Ghazaryan2018}%
  \BibitemOpen
  \bibfield  {author} {\bibinfo {author} {\bibfnamefont {D.}~\bibnamefont
  {Ghazaryan}}, \bibinfo {author} {\bibfnamefont {M.~T.}\ \bibnamefont
  {Greenaway}}, \bibinfo {author} {\bibfnamefont {Z.}~\bibnamefont {Wang}},
  \bibinfo {author} {\bibfnamefont {V.~H.}\ \bibnamefont {Guarochico-Moreira}},
  \bibinfo {author} {\bibfnamefont {I.~J.}\ \bibnamefont {Vera-Marun}},
  \bibinfo {author} {\bibfnamefont {J.}~\bibnamefont {Yin}}, \bibinfo {author}
  {\bibfnamefont {Y.}~\bibnamefont {Liao}}, \bibinfo {author} {\bibfnamefont
  {S.~V.}\ \bibnamefont {Morozov}}, \bibinfo {author} {\bibfnamefont
  {O.}~\bibnamefont {Kristanovski}}, \bibinfo {author} {\bibfnamefont {A.~I.}\
  \bibnamefont {Lichtenstein}}, \bibinfo {author} {\bibfnamefont {M.~I.}\
  \bibnamefont {Katsnelson}}, \bibinfo {author} {\bibfnamefont
  {F.}~\bibnamefont {Withers}}, \bibinfo {author} {\bibfnamefont
  {A.}~\bibnamefont {Mishchenko}}, \bibinfo {author} {\bibfnamefont
  {L.}~\bibnamefont {Eaves}}, \bibinfo {author} {\bibfnamefont {A.~K.}\
  \bibnamefont {Geim}}, \bibinfo {author} {\bibfnamefont {K.~S.}\ \bibnamefont
  {Novoselov}}, \ and\ \bibinfo {author} {\bibfnamefont {A.}~\bibnamefont
  {Misra}},\ }\href@noop {} {\bibfield  {journal} {\bibinfo  {journal} {Nat.
  Electron.}\ }\textbf {\bibinfo {volume} {1}},\ \bibinfo {pages} {344}
  (\bibinfo {year} {2018})}\BibitemShut {NoStop}%
\bibitem [{\citenamefont {Abramchuk}\ \emph {et~al.}(2018)\citenamefont
  {Abramchuk}, \citenamefont {Jaszewski}, \citenamefont {Metz}, \citenamefont
  {Osterhoudt}, \citenamefont {Wang}, \citenamefont {Burch},\ and\
  \citenamefont {Tafti}}]{Abramchuk2018}%
  \BibitemOpen
  \bibfield  {author} {\bibinfo {author} {\bibfnamefont {M.}~\bibnamefont
  {Abramchuk}}, \bibinfo {author} {\bibfnamefont {S.}~\bibnamefont
  {Jaszewski}}, \bibinfo {author} {\bibfnamefont {K.~R.}\ \bibnamefont {Metz}},
  \bibinfo {author} {\bibfnamefont {G.~B.}\ \bibnamefont {Osterhoudt}},
  \bibinfo {author} {\bibfnamefont {Y.}~\bibnamefont {Wang}}, \bibinfo {author}
  {\bibfnamefont {K.~S.}\ \bibnamefont {Burch}}, \ and\ \bibinfo {author}
  {\bibfnamefont {F.}~\bibnamefont {Tafti}},\ }\href@noop {} {\bibfield
  {journal} {\bibinfo  {journal} {Adv. Mater.}\ }\textbf {\bibinfo {volume}
  {30}},\ \bibinfo {pages} {1801325} (\bibinfo {year} {2018})}\BibitemShut
  {NoStop}%
\bibitem [{\citenamefont {McGuire}\ \emph {et~al.}(2017)\citenamefont
  {McGuire}, \citenamefont {Clark}, \citenamefont {Kc}, \citenamefont {Chance},
  \citenamefont {Jellison}, \citenamefont {Cooper}, \citenamefont {Xu},\ and\
  \citenamefont {Sales}}]{McGuire2017}%
  \BibitemOpen
  \bibfield  {author} {\bibinfo {author} {\bibfnamefont {M.~A.}\ \bibnamefont
  {McGuire}}, \bibinfo {author} {\bibfnamefont {G.}~\bibnamefont {Clark}},
  \bibinfo {author} {\bibfnamefont {S.}~\bibnamefont {Kc}}, \bibinfo {author}
  {\bibfnamefont {W.~M.}\ \bibnamefont {Chance}}, \bibinfo {author}
  {\bibfnamefont {G.~E.}\ \bibnamefont {Jellison}}, \bibinfo {author}
  {\bibfnamefont {V.~R.}\ \bibnamefont {Cooper}}, \bibinfo {author}
  {\bibfnamefont {X.}~\bibnamefont {Xu}}, \ and\ \bibinfo {author}
  {\bibfnamefont {B.~C.}\ \bibnamefont {Sales}},\ }\href@noop {} {\bibfield
  {journal} {\bibinfo  {journal} {Phys. Rev. Mater.}\ }\textbf {\bibinfo
  {volume} {1}},\ \bibinfo {pages} {014001} (\bibinfo {year}
  {2017})}\BibitemShut {NoStop}%
\bibitem [{\citenamefont {Gong}\ \emph {et~al.}(2017)\citenamefont {Gong},
  \citenamefont {Li}, \citenamefont {Li}, \citenamefont {Ji}, \citenamefont
  {Stern}, \citenamefont {Xia}, \citenamefont {Cao}, \citenamefont {Bao},
  \citenamefont {Wang}, \citenamefont {Wang}, \citenamefont {Qiu},
  \citenamefont {Cava}, \citenamefont {Louie}, \citenamefont {Xia},\ and\
  \citenamefont {Zhang}}]{Gong2017}%
  \BibitemOpen
  \bibfield  {author} {\bibinfo {author} {\bibfnamefont {C.}~\bibnamefont
  {Gong}}, \bibinfo {author} {\bibfnamefont {L.}~\bibnamefont {Li}}, \bibinfo
  {author} {\bibfnamefont {Z.}~\bibnamefont {Li}}, \bibinfo {author}
  {\bibfnamefont {H.}~\bibnamefont {Ji}}, \bibinfo {author} {\bibfnamefont
  {A.}~\bibnamefont {Stern}}, \bibinfo {author} {\bibfnamefont
  {Y.}~\bibnamefont {Xia}}, \bibinfo {author} {\bibfnamefont {T.}~\bibnamefont
  {Cao}}, \bibinfo {author} {\bibfnamefont {W.}~\bibnamefont {Bao}}, \bibinfo
  {author} {\bibfnamefont {C.}~\bibnamefont {Wang}}, \bibinfo {author}
  {\bibfnamefont {Y.}~\bibnamefont {Wang}}, \bibinfo {author} {\bibfnamefont
  {Z.~Q.}\ \bibnamefont {Qiu}}, \bibinfo {author} {\bibfnamefont {R.~J.}\
  \bibnamefont {Cava}}, \bibinfo {author} {\bibfnamefont {S.~G.}\ \bibnamefont
  {Louie}}, \bibinfo {author} {\bibfnamefont {J.}~\bibnamefont {Xia}}, \ and\
  \bibinfo {author} {\bibfnamefont {X.}~\bibnamefont {Zhang}},\ }\href@noop {}
  {\bibfield  {journal} {\bibinfo  {journal} {Nature}\ }\textbf {\bibinfo
  {volume} {546}},\ \bibinfo {pages} {265} (\bibinfo {year}
  {2017})}\BibitemShut {NoStop}%
\bibitem [{\citenamefont {Dai}(2015)}]{dai2015antiferromagnetic}%
  \BibitemOpen
  \bibfield  {author} {\bibinfo {author} {\bibfnamefont {P.}~\bibnamefont
  {Dai}},\ }\href@noop {} {\bibfield  {journal} {\bibinfo  {journal} {Rev. Mod.
  Phys.}\ }\textbf {\bibinfo {volume} {87}},\ \bibinfo {pages} {855} (\bibinfo
  {year} {2015})}\BibitemShut {NoStop}%
\bibitem [{\citenamefont {Pientka}\ \emph {et~al.}(2017)\citenamefont
  {Pientka}, \citenamefont {Keselman}, \citenamefont {Berg}, \citenamefont
  {Yacoby}, \citenamefont {Stern},\ and\ \citenamefont
  {Halperin}}]{Pientka2017}%
  \BibitemOpen
  \bibfield  {author} {\bibinfo {author} {\bibfnamefont {F.}~\bibnamefont
  {Pientka}}, \bibinfo {author} {\bibfnamefont {A.}~\bibnamefont {Keselman}},
  \bibinfo {author} {\bibfnamefont {E.}~\bibnamefont {Berg}}, \bibinfo {author}
  {\bibfnamefont {A.}~\bibnamefont {Yacoby}}, \bibinfo {author} {\bibfnamefont
  {A.}~\bibnamefont {Stern}}, \ and\ \bibinfo {author} {\bibfnamefont {B.~I.}\
  \bibnamefont {Halperin}},\ }\href@noop {} {\bibfield  {journal} {\bibinfo
  {journal} {Phys. Rev. X}\ }\textbf {\bibinfo {volume} {7}},\ \bibinfo {pages}
  {021032} (\bibinfo {year} {2017})}\BibitemShut {NoStop}%
\bibitem [{\citenamefont {Ren}\ \emph {et~al.}(2019)\citenamefont {Ren},
  \citenamefont {Pientka}, \citenamefont {Hart}, \citenamefont {Pierce},
  \citenamefont {Kosowsky}, \citenamefont {Lunczer}, \citenamefont {Schlereth},
  \citenamefont {Scharf}, \citenamefont {Hankiewicz}, \citenamefont
  {Molenkamp}, \citenamefont {Halperin},\ and\ \citenamefont
  {Yacoby}}]{RenHC2019}%
  \BibitemOpen
  \bibfield  {author} {\bibinfo {author} {\bibfnamefont {H.}~\bibnamefont
  {Ren}}, \bibinfo {author} {\bibfnamefont {F.}~\bibnamefont {Pientka}},
  \bibinfo {author} {\bibfnamefont {S.}~\bibnamefont {Hart}}, \bibinfo {author}
  {\bibfnamefont {A.~T.}\ \bibnamefont {Pierce}}, \bibinfo {author}
  {\bibfnamefont {M.}~\bibnamefont {Kosowsky}}, \bibinfo {author}
  {\bibfnamefont {L.}~\bibnamefont {Lunczer}}, \bibinfo {author} {\bibfnamefont
  {R.}~\bibnamefont {Schlereth}}, \bibinfo {author} {\bibfnamefont
  {B.}~\bibnamefont {Scharf}}, \bibinfo {author} {\bibfnamefont {E.~M.}\
  \bibnamefont {Hankiewicz}}, \bibinfo {author} {\bibfnamefont {L.~W.}\
  \bibnamefont {Molenkamp}}, \bibinfo {author} {\bibfnamefont {B.~I.}\
  \bibnamefont {Halperin}}, \ and\ \bibinfo {author} {\bibfnamefont
  {A.}~\bibnamefont {Yacoby}},\ }\href@noop {} {\bibfield  {journal} {\bibinfo
  {journal} {Nature}\ }\textbf {\bibinfo {volume} {569}},\ \bibinfo {pages}
  {93} (\bibinfo {year} {2019})}\BibitemShut {NoStop}%
\bibitem [{\citenamefont {Fornieri}\ \emph {et~al.}(2019)\citenamefont
  {Fornieri}, \citenamefont {Whiticar}, \citenamefont {Setiawan}, \citenamefont
  {Portol¨¦s}, \citenamefont {Drachmann}, \citenamefont {Keselman},
  \citenamefont {Gronin}, \citenamefont {Thomas}, \citenamefont {Wang},
  \citenamefont {Kallaher}, \citenamefont {Gardner}, \citenamefont {Berg},
  \citenamefont {Manfra}, \citenamefont {Stern}, \citenamefont {Marcus},\ and\
  \citenamefont {Nichele}}]{FornieriA2019}%
  \BibitemOpen
  \bibfield  {author} {\bibinfo {author} {\bibfnamefont {A.}~\bibnamefont
  {Fornieri}}, \bibinfo {author} {\bibfnamefont {A.~M.}\ \bibnamefont
  {Whiticar}}, \bibinfo {author} {\bibfnamefont {F.}~\bibnamefont {Setiawan}},
  \bibinfo {author} {\bibfnamefont {E.}~\bibnamefont {Portol¨¦s}}, \bibinfo
  {author} {\bibfnamefont {A.~C.~C.}\ \bibnamefont {Drachmann}}, \bibinfo
  {author} {\bibfnamefont {A.}~\bibnamefont {Keselman}}, \bibinfo {author}
  {\bibfnamefont {S.}~\bibnamefont {Gronin}}, \bibinfo {author} {\bibfnamefont
  {C.}~\bibnamefont {Thomas}}, \bibinfo {author} {\bibfnamefont
  {T.}~\bibnamefont {Wang}}, \bibinfo {author} {\bibfnamefont {R.}~\bibnamefont
  {Kallaher}}, \bibinfo {author} {\bibfnamefont {G.~C.}\ \bibnamefont
  {Gardner}}, \bibinfo {author} {\bibfnamefont {E.}~\bibnamefont {Berg}},
  \bibinfo {author} {\bibfnamefont {M.~J.}\ \bibnamefont {Manfra}}, \bibinfo
  {author} {\bibfnamefont {A.}~\bibnamefont {Stern}}, \bibinfo {author}
  {\bibfnamefont {C.~M.}\ \bibnamefont {Marcus}}, \ and\ \bibinfo {author}
  {\bibfnamefont {F.}~\bibnamefont {Nichele}},\ }\href@noop {} {\bibfield
  {journal} {\bibinfo  {journal} {Nature}\ }\textbf {\bibinfo {volume} {569}},\
  \bibinfo {pages} {89} (\bibinfo {year} {2019})}\BibitemShut {NoStop}%
\bibitem [{\citenamefont {{Pahomi}}\ \emph {et~al.}(2019)\citenamefont
  {{Pahomi}}, \citenamefont {{Sigrist}},\ and\ \citenamefont
  {{Soluyanov}}}]{Pahomi2019}%
  \BibitemOpen
  \bibfield  {author} {\bibinfo {author} {\bibfnamefont {T.~E.}\ \bibnamefont
  {{Pahomi}}}, \bibinfo {author} {\bibfnamefont {M.}~\bibnamefont {{Sigrist}}},
  \ and\ \bibinfo {author} {\bibfnamefont {A.~A.}\ \bibnamefont
  {{Soluyanov}}},\ }\href@noop {} {\bibfield  {journal} {\bibinfo  {journal}
  {arXiv:1904.07822}\ } (\bibinfo {year} {2019})}\BibitemShut {NoStop}%
\bibitem [{\citenamefont {Zhang}\ \emph
  {et~al.}(2019{\natexlab{c}})\citenamefont {Zhang}, \citenamefont {Cole},
  \citenamefont {Wu},\ and\ \citenamefont {Das~Sarma}}]{ZhangRX2019}%
  \BibitemOpen
  \bibfield  {author} {\bibinfo {author} {\bibfnamefont {R.-X.}\ \bibnamefont
  {Zhang}}, \bibinfo {author} {\bibfnamefont {W.~S.}\ \bibnamefont {Cole}},
  \bibinfo {author} {\bibfnamefont {X.}~\bibnamefont {Wu}}, \ and\ \bibinfo
  {author} {\bibfnamefont {S.}~\bibnamefont {Das~Sarma}},\ }\href@noop {}
  {\bibfield  {journal} {\bibinfo  {journal} {Phys. Rev. Lett.}\ }\textbf
  {\bibinfo {volume} {123}},\ \bibinfo {pages} {167001} (\bibinfo {year}
  {2019}{\natexlab{c}})}\BibitemShut {NoStop}%
\bibitem [{\citenamefont {Wu}\ \emph {et~al.}(2019{\natexlab{b}})\citenamefont
  {Wu}, \citenamefont {Hou}, \citenamefont {Luo}, \citenamefont {Li},\ and\
  \citenamefont {Zhang}}]{wu2019inplane}%
  \BibitemOpen
  \bibfield  {author} {\bibinfo {author} {\bibfnamefont {Y.-J.}\ \bibnamefont
  {Wu}}, \bibinfo {author} {\bibfnamefont {J.}~\bibnamefont {Hou}}, \bibinfo
  {author} {\bibfnamefont {X.}~\bibnamefont {Luo}}, \bibinfo {author}
  {\bibfnamefont {Y.}~\bibnamefont {Li}}, \ and\ \bibinfo {author}
  {\bibfnamefont {C.}~\bibnamefont {Zhang}},\ }\href@noop {} {\bibfield
  {journal} {\bibinfo  {journal} {arXiv:1905.08896}\ } (\bibinfo {year}
  {2019}{\natexlab{b}})}\BibitemShut {NoStop}%
\end{thebibliography}%


\end{document}